\documentclass[twocolumn]{aastex631}

\usepackage{amsmath, color, natbib,graphicx,tablefootnote,subfigure}
\usepackage{graphics,subfigure,hyperref,natbib,float,appendix}
\usepackage{apjfonts}

\newcommand{\ltsima}{$\; \buildrel < \over \sim \;$}
\newcommand{\simlt}{\lower.5ex\hbox{\ltsima}}

\def\arcdeg{\hbox{$^\circ$}}

\def\arcsec{\hbox{$^{\prime\prime}$}}

\newcommand\iv{{\sc iv}}
\newcommand\W{{$\lambda$}}
\newcommand{\CH}[1]{\colhead{#1}}
\newcommand\hst{{{\it HST}}}

\newcommand{\aref}[1]{\hyperref[#1]{Appendix~\ref{#1}}}

\defcitealias{berg16}{B16}
\defcitealias{berg19a}{B19a}
\defcitealias{berg19b}{B19b}
\defcitealias{berg21}{Paper {\sc i}}

\submitjournal{AASJournal ApJ}
\shortauthors{Olivier et al.}
\shorttitle{EELGs}

\begin{document}
\title{Characterizing Extreme Emission Line Galaxies II: \\
A Self-Consistent Model of Their Ionizing Spectrum\footnote{
Based on observations made with the NASA/ESA Hubble Space Telescope,
obtained from the Data Archive at the Space Telescope Science Institute, which
is operated by the Association of Universities for Research in Astronomy, Inc.,
under NASA contract NASA 5-26555.}}

\author[0000-0002-4606-4240]{Grace M. Olivier}
\affiliation{Department of Astronomy, The Ohio State University, 140 W. 18th Ave., Columbus, Ohio 43210, USA}
\affiliation{Center for Cosmology \& AstroParticle Physics, The Ohio State University, 191 W Woodruff Avenue, Columbus, OH 43210}
\email{olivier.15@osu.edu}

\author[0000-0002-4153-053X]{Danielle A. Berg}
\affiliation{Department of Astronomy, The University of Texas at Austin, 2515 Speedway, Stop C1400, Austin, TX 78712, USA}

\author[0000-0002-0302-2577]{John Chisholm}
\affiliation{Department of Astronomy, The University of Texas at Austin, 2515 Speedway, Stop C1400, Austin, TX 78712, USA}

\author[0000-0001-9714-2758]{Dawn K. Erb}
\affiliation{The Leonard E. Parker Center for Gravitation, Cosmology and Astrophysics, Department of Physics, University of Wisconsin-Milwaukee, 3135 North Maryland Avenue, Milwaukee, WI 53211, USA}

\author[0000-0003-1435-3053]{Richard W. Pogge}
\affiliation{Department of Astronomy, The Ohio State University, 140 W. 18th Ave., Columbus, Ohio 43210, USA}
\affiliation{Center for Cosmology \& AstroParticle Physics, The Ohio State University, 191 W Woodruff Avenue, Columbus, OH 43210}

\author[0000-0003-0605-8732]{Evan D. Skillman}
\affiliation{Minnesota Institute for Astrophysics, University of Minnesota, 116 Church Street SE, Minneapolis, MN 55455, USA}


\begin{abstract}

Observations of high-redshift galaxies ($z >$ 5) have shown that these galaxies have
extreme emission lines with equivalent widths much larger than their local
star-forming counterparts.
Extreme emission line galaxies (EELGs) in the nearby universe are likely analogues
to galaxies during the Epoch of Reionization and provide nearby laboratories
to understand the physical processes important to the early universe.
We use \textit{HST}/COS and LBT/MODS spectra to study two nearby EELGs, J104457
and J141851.
The FUV spectra indicate that these two galaxies contain stellar populations
with ages \simlt\ 10 Myr and metallicities $\leq$0.15 Z$_\odot$.
We use photoionization modeling to compare emission lines from models of 
single-age bursts of star-formation to observed emission lines and find
that the single-age bursts do not reproduce high-ionization lines including
[\ion{O}{3}] or very-high ionization lines like \ion{He}{2} or [\ion{O}{4}].
Photoionization modeling using the stellar populations fit from the UV continuum
similarly are not capable of reproducing the emission
lines from the very-high ionization zone.
We add a blackbody to the stellar populations fit from the UV continuum to
model the necessary high-energy photons to reproduce the very-high ionization
lines of \ion{He}{2} and [\ion{O}{4}].
We find that we need a blackbody of 80,000 K and $\sim$60-70\% of the luminosity
from the young stellar population to reproduce the very-high ionization lines
while simultaneously reproducing the low-, intermediate-, and high-ionization
emission lines.
Our self-consistent model of the ionizing spectra of two nearby EELGs 
indicates the presence of a previously unaccounted-for source of hard
ionizing photons in reionization analogues.

\end{abstract} 

\keywords{Dwarf galaxies (416), Ultraviolet astronomy (1736), Galaxy chemical evolution (580), Galaxy spectroscopy (2171), High-redshift galaxies (734), Emission line galaxies (459)}


\section{Introduction}

The first galaxies are likely responsible for producing the photons that 
ionized the universe during the Epoch of Reionization 
\citep[EoR,][]{wise14,robertson15,madau15,stanway16}.
With telescopes like the \textit{Hubble Space Telescope} (\textit{HST}),
Atacama Large Millimeter/submillimeter Array (ALMA), and other telescopes, 
astronomers have discovered thousands of galaxies in the high-redshift 
universe ($z >$ 6) in an attempt to study these first galaxies
\citep[e.g.,][]{bouwens15,finkelstein15,livermore17,atek18,oesch18,bouwens20,harikane20}.
More direct observations of these first galaxies and their environments 
during the EoR are just on the horizon with the coming 
era of the \textit{James Webb Space Telescope} (\textit{JWST}) and 
extremely large telescopes (ELTs), in combination with next-generation 
\ion{H}{1} surveys such as the Square Kilometer Array (SKA).
Soon we will be able to observe the first galaxies and their effects
on the neutral gas surrounding them.

Over the past decade, the rest-frame UV emission-line spectra have been 
observed from several very high redshift galaxies ($z > 6$) revealing 
strong, high-ionization nebular lines \citep[e.g.,][]{stark15,stark16,mainali18} 
that are indicative of very hard radiation fields in low-metallicity 
environments.
With this in mind, much can be learned from detailed studies of nearby
galaxies with properties analogous to what we expect to find in the distant universe. 
In particular, recent studies have focused on metal-poor, high-ionization 
galaxies with high specific star formation rates both locally 
\citep[e.g.,][]{berg16,izotov16,jaskot17,senchyna17,izotov18, senchyna20,berg19a,berg19b} 
and at intermediate redshifts using lensed galaxies 
\citep[e.g.,][]{hainline09,bayliss14,christensen14,james14,rigby18a,berg18}.

These recently discovered samples of Reionization Analogues have
extreme equivalent widths (EWs) in their emission lines from high-ionization 
species like \ion{O}{3}], \ion{C}{3}], \ion{C}{4}, and \ion{He}{2}.
As a result of their large equivalent widths and hard radiation fields, 
extreme emission line galaxies (EELGs) in the local universe provide a 
laboratory for understanding the physical processes active in the
first galaxies.
The large EWs from \ion{O}{3}], \ion{C}{4}, and \ion{He}{2} in galaxies
out to $z \sim$ 2 have been associated with low metallicity environments 
and very young stellar populations from recent bursts of star-formation
producing highly ionized gas
\citep{atek11,vanderwel11,maseda13,maseda14,rigby15,berg16,senchyna17,chevallard18,berg19a,berg19b,tang19,tang21}.
Exploring similarly low metallicity environments in local EELGs can
help us understand those young stellar populations and the highly
ionized gas.

These EELGs often demonstrate strong \ion{He}{2} emission along with
their other extreme emission lines, but the mechanisms producing the 
high-energy photons at $>$54eV are rarely understood.
A number of studies have explored the production mechanism for the extreme 
\ion{He}{2} emission at $\lambda$1640 in the restframe UV and $\lambda$4686 
in the optical in a number of these reionization analogues
\citep[e.g.,][]{schaerer02, kehrig15,kehrig18,kehrig21,schaerer19,senchyna17,senchyna20}.
Some of the postulated sources of the \ion{He}{2} emission include:
Wolf-Rayet stars at low metallicity, high-mass X-ray binaries, 
metal-poor massive stars, and shocks, but, when examined in more detail,
these suggestions tend to be inadequate in their photon production in the
energy range needed for extreme \ion{He}{2} emission \citep{kehrig15,kehrig18,senchyna20,wofford21}.

Here we examine two of these reionization era analogue galaxies.
The first of these galaxies is J104457 which has the largest \ion{C}{4} $\lambda \lambda$1548,1550 \AA\
EW (-6.71\AA\ ,-2.83\AA\ ) measured in the local universe \citep{berg16}.
The second galaxy is J141851 which has the largest \ion{He}{2} $\lambda$1640 \AA\  
EW (-2.82\AA\ ) measured in the local universe \citep{berg19a}.
These two galaxies are prime targets for studying the physical processes required to create
high ionization gas and to explore the conditions present in the EoR.
A first look at the high-energy ionizing photons in these galaxies was presented in \cite{berg19b},
but these fascinating objects demand a more complete analysis.
Here we present the second paper in a series of work to characterize extreme emission line galaxies.
In Paper {\sc i}, \citet{berg21}, we presented a detailed examination of the extreme emission lines, 
nebular physical conditions, and chemical abundances in J104457 and J141851.
Owing to the detection of a number of very high-ionization emission lines 
(e.g., \ion{C}{4}, \ion{He}{2}, \ion{O}{4}, \ion{Si}{4}, \ion{Ne}{5}, and \ion{Ar}{4}), 
we adopted a novel 4-zone ionization model and found both EELGs to be characterized 
by much higher ionization parameters than reported in previous works.
The main result of \citetalias{berg21} highlights a persistent 
{\it high energy ionizing photon production problem} (HEIP$^3$) that motivates 
the self-consistent modeling of their ionizing spectra presented here in Paper {\sc ii}.

Paper {\sc ii} is structured as follows:
We present the optical and UV spectra of J104457 and J141851 in \autoref{sec:data}.
In \autoref{sec:continuum}, we investigate the UV stellar continuum.
The procedure used for the stellar continuum fits is discussed in \S~\ref{sec:contmethods}
and the resulting stellar population properties are presented in \S~\ref{sec:spops}.
Additionally, we generalize our findings in order to characterize the spectra of 
extremely metal poor stellar populations in \S~\ref{sec:specfeat}.
\autoref{sec:photoion} then combines constraints from the stellar continuum fits of
\autoref{sec:continuum} with the emission line ratios from the optical and UV spectra
using \texttt{CLOUDY} \citep{ferland17} photoionization modeling.
We explore several input ionizing sources in \autoref{sec:photmods}, including
single-aged burst binary models (\S~\ref{sec:mod1},
the best-fit stellar population models (\S~\ref{sec:mod2}),
and combined stellar plus blackbody models in (\S~\ref{sec:mod3}).
The resulting best fit to the suite of observed emission lines is produced by the latter models,
requiring an additional source of very-high-energy ionizing photons.
We discuss this result and the possible physical origin of the very-high-energy photons in \autoref{sec:discuss}.
Finally, we present our conclusions in \autoref{sec:conclusions}.


\begin{deluxetable}{rlcc}
\tablecaption{Extreme UV Emission-Line Galaxy Properties}
\tablehead{
\multicolumn{1}{r}{Property}  & \multicolumn{1}{l}{Units}	& \CH{J104457} 		& \CH{J141851}  }
\startdata	
\multicolumn{4}{c}{\bf Adopted from Archival Sources:} \\
Reference		&					    & \citetalias{berg16}       & \citetalias{berg19a} \\
RA              &                       & 10:44:57.79 & 14:18:51.13 \\
Dec             &                       & +03:53:13.15 & +21:02:39.74 \\
$z$				&					    & 0.013	                    & 0.009         	\\
log $M_\star$	& M$_\odot$			    & 6.80		                & 6.63          	\\
log SFR			& M$_\odot$ yr$^{-1}$	& $-0.85$		            & $-1.16$       	\\
log sSFR		& yr$^{-1}$			    & $-7.64$		            & $-7.79$         	\\
$E_(B-V)$		& mag.				    & 0.077				        & 0.140		    	\\
$12+$log(O/H)	& dex ($Z_\odot$)	    & $7.45$ ($0.058$) 	        & $7.54$ ($0.071$)	\\
log $U$			&					    & $-1.77$				    & $-2.42$			\\
\\
\multicolumn{4}{c}{\bf Derived from the UV COS G160M Spectra:} \\
EW$_{\rm OIII]}$		& \AA\		    & $-2.89,-6.17$				& $-1.68,-4.78$		\\
EW$_{\rm CIV}$		    & \AA\			& $-6.71,-2.83$				& $-1.78,-1.43$		\\
EW$_{\rm HeII}$		    & \AA\			& $-2.75$					& $-2.82$			\\
\ion{O}{3}] $\Delta V$ 	& km s$^{-1}$	& $87.3$					& $64.7$			\\
\ion{C}{4} $\Delta V$	& km s$^{-1}$	& $106.7$, $86.2$			& $101.8$, $78.6$  \\
\hline
COS FWHM            & km s$^{-1}$       & 188.4                     & 103.5 \\
\enddata
\tablecomments{ 
Properties of the extreme UV emission-line galaxies presented here.
The top portion of the table lists properties previously reported by
\citet{berg16} for J104457 and \citet{berg19a} for J141851.
The R.A., Decl., redshift, total stellar masses, SFRs, and sSFRs were adopted from the SDSS MPA-JHU DR8 
catalog\footnote{Data catalogues are available from \url{http://www.sdss3.org/dr10/spectro/galaxy_mpajhu.php}.
The Max Planck Institute for Astrophysics/John Hopkins University(MPA/JHU) SDSS data base was produced by a 
collaboration of researchers(currently or formerly) from the MPA and the JHU. 
The team is made up of Stephane Charlot (IAP), Guinevere Kauffmann and Simon White (MPA),
Tim Heckman (JHU), Christy Tremonti (U. Wisconsin-Madison $-$ formerly JHU) and Jarle 
Brinchmann (Leiden University $-$ formerly MPA).},
whereas $E(B-V)$, $12+$log(O/H), and ionization parameter log $U$, 
were measured from the SDSS optical spectra.
The bottom portion of the table lists the properties derived in \cite{berg19b} from the UV
HST/COS G160M spectra (see Figure~2).
Equivalent widths are listed for \ion{C}{4} \W\W1548,1550, \ion{O}{3} \W\W1661,1666, and \ion{He}{2} \W1640.
We list velocity widths (FWHM), $\Delta V$, of \ion{O}{3}] \W1666 and \ion{He}{2} \W1640,
and both blue and red components of the \ion{C}{4} \W1548 profiles.
Since the individual \ion{C}{4} lines present as doublets, we list the 
$V_{peak}^{blue}$, $V_{peak}^{red}$, and $V_{sep.}$ measured for the \ion{C}{4} \W1548 line. 
The final line of the table lists the spectral resolution from COS used in our continuum fitting.
\looseness=-2}
\label{tbl1}
\end{deluxetable}


\section{Extreme UV Emission-Line Galaxy Observations} \label{sec:data}


The main objective of this work is to investigate whether the suite of nebular 
emission lines from two EELGs can be self-consistently reproduced by photoionization
models using the ionizing photons from the massive star populations.
To analyze the interplay of the ionizing stellar population and nebular
gas of J104457 and J141851, we first measure the nebular emission lines from their
UV and optical spectra.


\subsection{\hst/COS UV Spectra}

In this work we used FUV spectra taken with the Cosmic Origins Spectrograph (COS)
on the \textit{Hubble Space Telescope} (\hst) that were first reported in 
\citet{berg16}, \citet{berg19a}, and \citet{berg19b} from the Hubble Space Telescope Project ID 15465.
We briefly summarize the observational details below.

All observations were centered on the NUV flux of J104457 and J141851 using the ACQ/Image acquisition mode.
Science exposures were then taken in the TIME-TAG mode using the 2.5\arcsec\ PSA aperture.
The FP-POS = ALL setting was used to take four images offset from one another in 
the dispersion direction, increasing the cumulative S/N and mitigating the effects 
of fixed pattern noise. 
Spectra were processed with CALCOS version 3.3.4, using the standard {\sc ExtractSegmentBoxcar}
aperture method, and binned by the 6 native COS pixels per resolution element such that 
the nominal $\delta v$ = 13.1 km s$^{-1}$.

The G160M spectra used a central wavelength of 1589 \AA\ and had total 
exposures of 6439 and 12374s for J104457 and J141851, respectively. 
This setup provides wavelength coverage that is rich with stellar and nebular features.
In particular, these spectra cover several important FUV high-ionization emission lines:
\ion{Si}{4} \W\W1394,1403, \ion{O}{4} \W\W1401,1405,1407, \ion{S}{4} \W\W1405,1406,1417,1424,
\ion{N}{4} \W\W1483,1487, \ion{C}{4} \W\W1548,1550, \ion{He}{2} \W1640, and \ion{O}{3}] \W\W1661,1666.
For the purpose of testing the production of nebular emission lines, not only are 
the \ion{Si}{4}, \ion{C}{4}, and \ion{He}{2} features difficult to interpret due to their potentially 
complex nebular and stellar contributions, but \ion{Si}{4} and \ion{C}{4} are also resonant lines.
Further, the \ion{S}{4} and \ion{N}{4} lines are rather faint or not detected in the UV spectra
of J104457 and J141851, and thus, only the \ion{O}{3}] and \ion{O}{4} lines, along with \ion{He}{2}, since its 
velocity profile shows the emission is mostly nebular in origin, remain to robustly constrain our models.
Therefore, in order to expand our suite of strong UV emission lines constraining the shape of the ionizing 
continuum, we also consider the low-resolution \hst/COS G140L spectra for these galaxies,
which provide access to the \ion{C}{3}] \W\W1907,1909 lines.

\begin{figure*}
\begin{center}
    \includegraphics[scale=0.5,trim=0mm 13mm 0mm 0mm,clip]{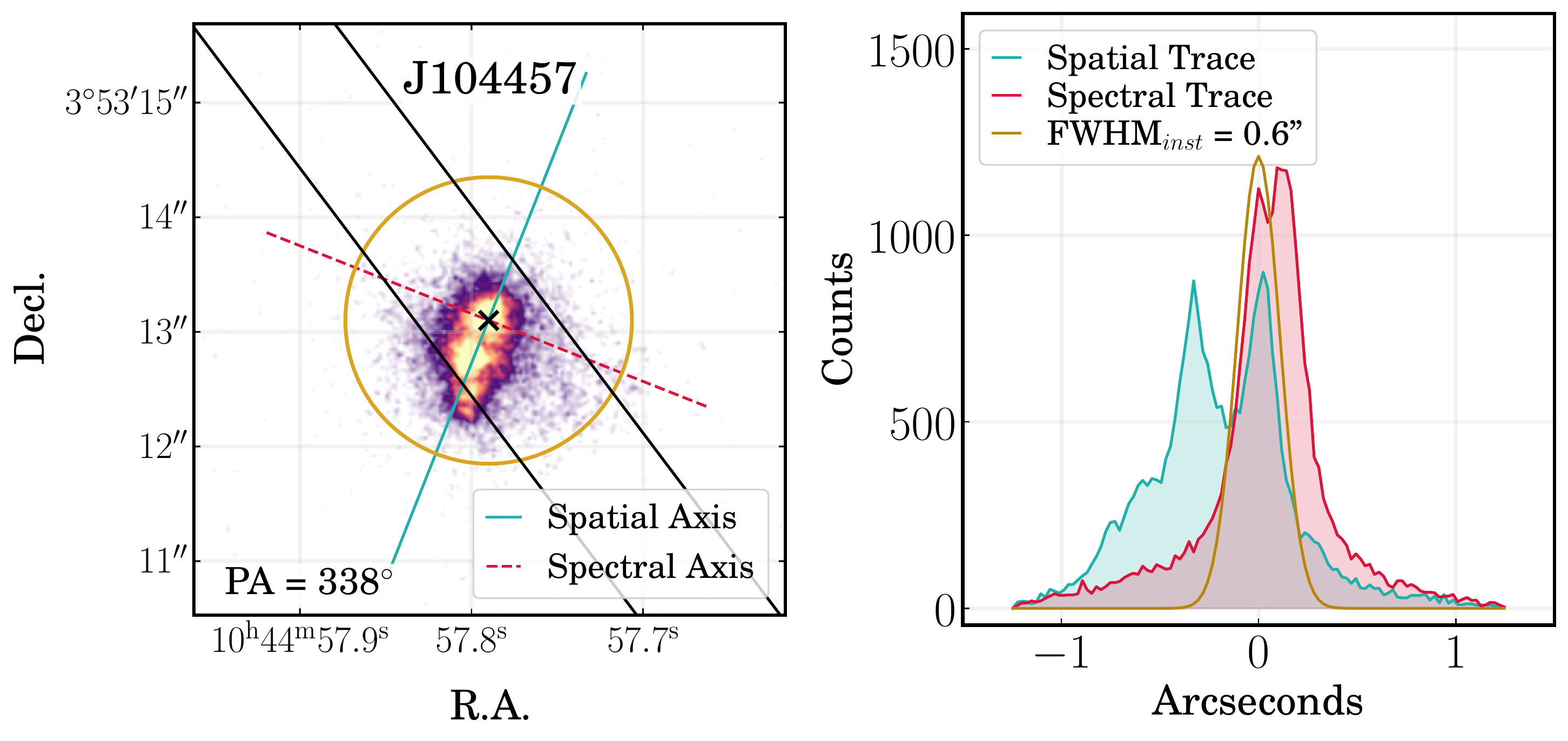} \\
    \includegraphics[scale=0.5,trim=0mm 2mm 0mm 0mm,clip]{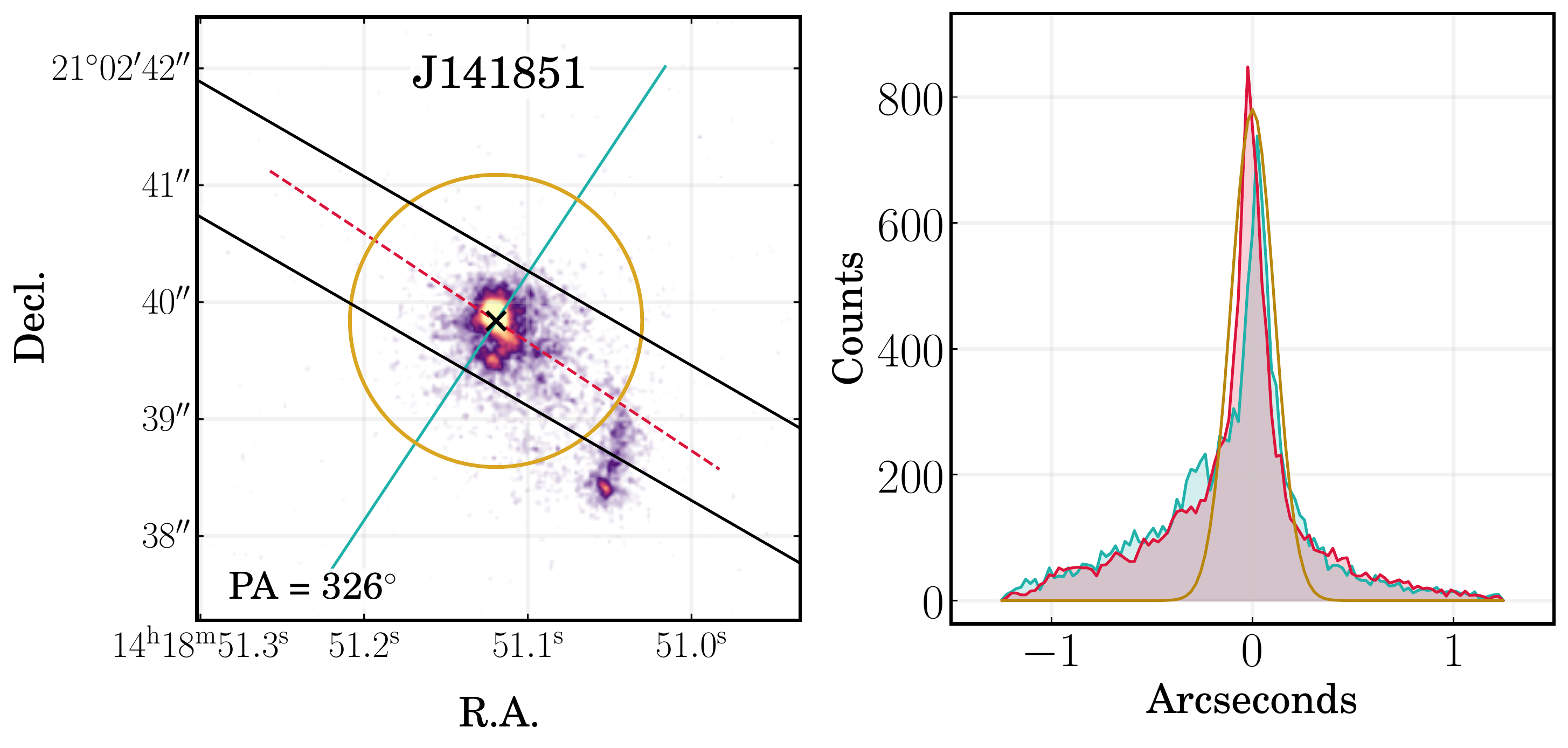} 
\end{center}
\caption{
The {\it HST}/COS NUV acquisition images of J104457 ({\it top}) and J141851 ({\it bottom}) 
are shown in the left hand column.
The 2.5\arcsec\ COS aperture used for the UV spectra is shown as a gold circle. 
The spectrograph position angle 
is noted in the corner and is used to determine the spatial and spectral axes.
In comparison, the 1\arcsec\ LBT/MODS slit (black lines) captures most of the NUV light, but
may miss a significant fraction of the extended nebular emission.
In the right hand column we plot the spatial and spectral light profiles captured within the COS aperture.
The 0.6\arcsec\ instrumental profile of COS is shown in gold.}
\label{fig1}
\end{figure*}

The NUV acquisition images, whose flux is dominated by the light from the young ionizing
stellar clusters, are shown in the left panels of \autoref{fig1}, and highlight the simple, 
very compact nature of these galaxies. 
The UV light from J104457 and J141851 is easily contained within the COS 2.5\arcsec\ aperture (gold circle),
corresponding to a physical size of $\lesssim 0.5$ kpc at their redshifts of $z\sim0.01$.
In the right panels of \autoref{fig1}, we plot the spatial light distribution perpendicular
to the dispersion axis (red dashed line in acquisition images).
The spatial trace is smoothed by a Gaussian 1D kernel with $\sigma=1$ pixel and 
compared to a Gaussian approximation of the full width half maximum (FWHM) of the non-Gaussian instrumental profile of 
FWHM$_{inst}=0.6$\arcsec\ (gold line). 
Similarly, the spectral light trace is plotted in green.
Comparing the observed spatial trace to the Gaussian profile estimates the spatial extent of each galaxy. 
J141851 has less extended emission and has a fitted FWHM$=0.12$\arcsec.
However, the spatial profile of J104457 is somewhat larger (FWHM$=0.7$\arcsec),
resulting in a degradation of the spectral resolution.

We measured the resolution of our spectral observations by fitting Gaussian profiles to 
three Milky Way absorption lines in the spectra of each galaxy 
(\ion{Si}{2} \W1526, \ion{C}{4} \W\W1548,1550, and \ion{Al}{2} \W1671).
For each galaxy we averaged the FWHM of the three lines together to find the 
spectral resolution. 
For J104457 and J141851, we found that the FWHM related to the spectral resolution was 188 km s$^{-1}$ and 104 km s$^{-1}$, respectively.
These values are larger than the native COS resolution:
nominally, the spectral resolution for the G160M grating is $R\sim20,000$ or $\Delta v = 15$ km s$^{-1}$,
with the observed spectral resolution broadened due to the considerable 
spatial extent of the galaxies (\autoref{fig1}). 
Similarly, \citet{berg19b} report GFWHMs of the \ion{O}{3}] nebular emission lines of 
$\Delta v = 87.3, 64.7$ km s$^{-1}$ for J104457 and J141851, respectively \citep{berg19b}.
Some of this velocity width could be due to the physical properties of the gas, 
but a degradation of the spectral resolution is also present.
Nonetheless, at these resolutions, the individual emission line fluxes and velocities of interest can
be resolved and precisely measured.


\subsection{Optical LBT/MODS Spectra}

While the FUV spectra enable modeling of the massive star populations,
the FUV contains a limited number of nebular emission lines of only a handful of ions.
Recently, \citetalias{berg21} presented a more detailed examination of the extreme emission lines in
J104457 and J141851 and their chemical abundances, using the \hst/COS G160M spectra together
with new optical spectra from the Multi-Object Double Spectrographs \citep[MODS,][]{pogge10} 
on the Large Binocular Telescope \citep[LBT,][]{hill10}.
We refer the reader to \citetalias{berg21} for a full description of the observations and \citet{berg15}
for a detailed discussion of the MODS data reduction pipeline.
For convenience, we present a brief overview below.

\begin{figure*}
\begin{center}
    \includegraphics[scale=0.225,trim=0mm 0mm 0mm 0mm,clip]{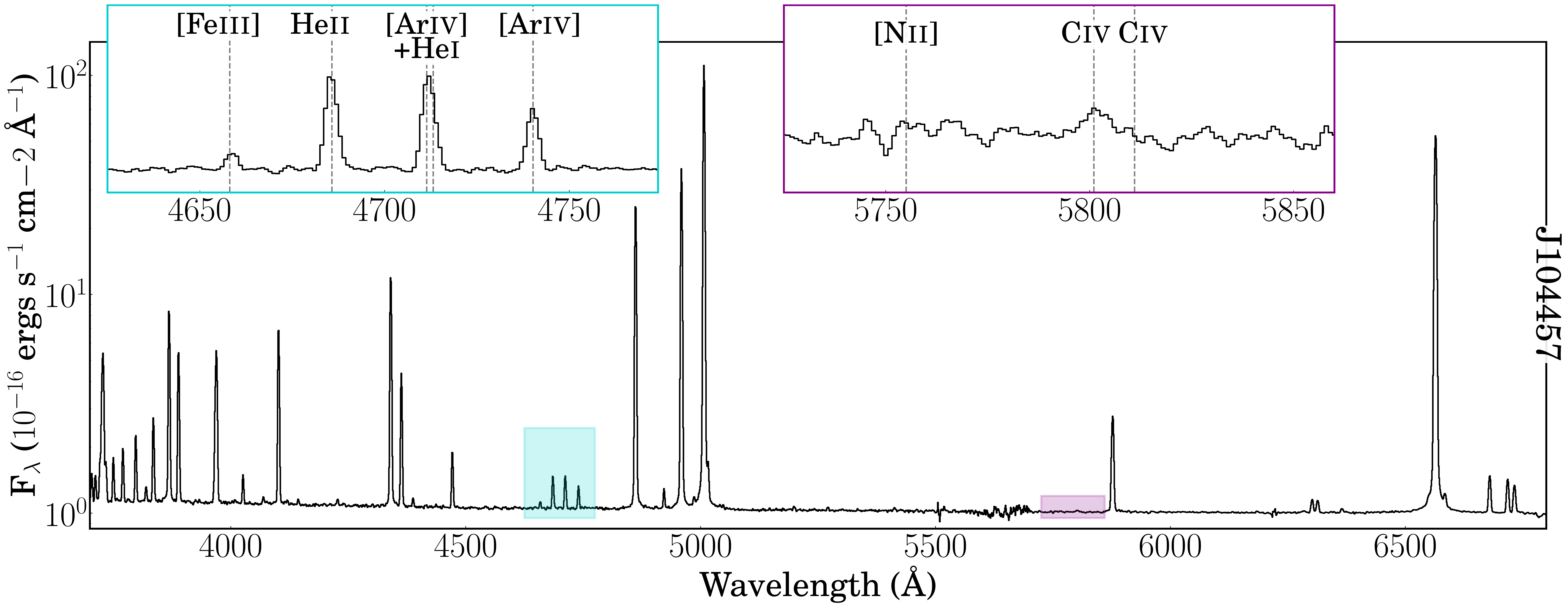}
    \includegraphics[scale=0.225,trim=0mm 0mm 0mm 0mm,clip]{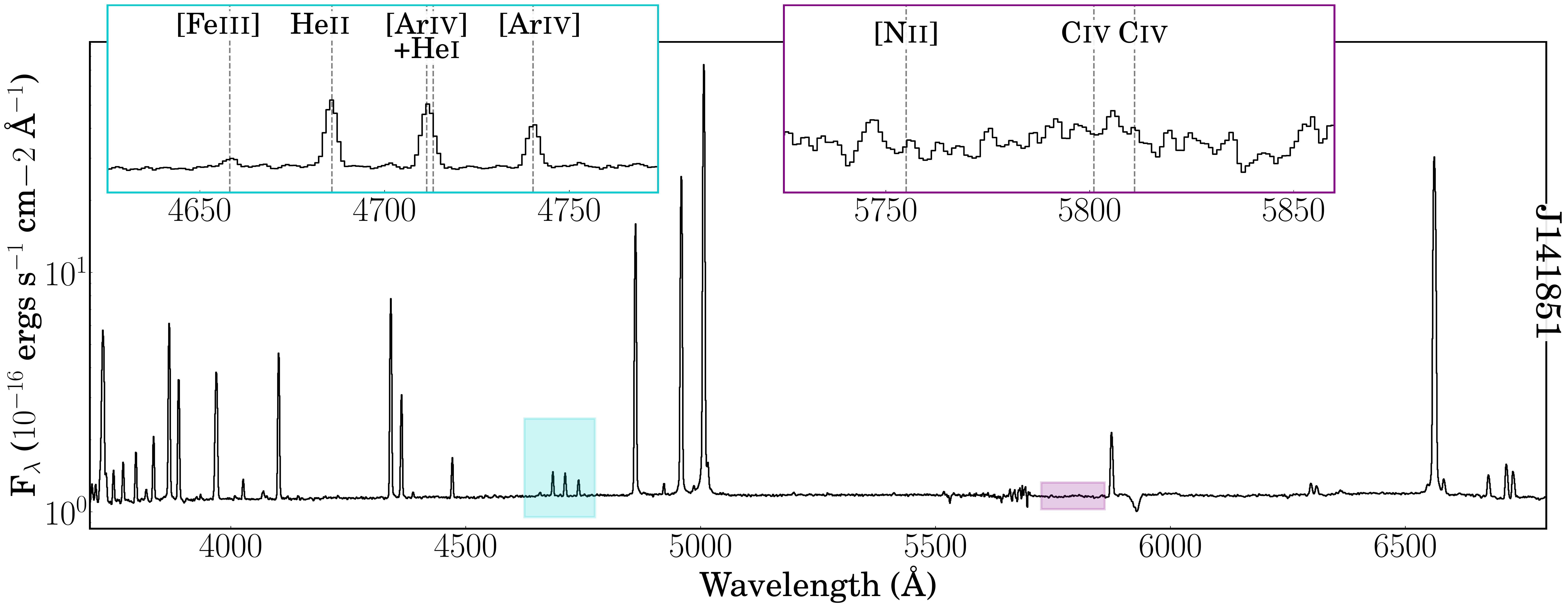}
\caption{
Optical LBT/MODS spectrum of J104457 ({\it top}) and J141851 ({\it bottom}), 
highlighting the blue and Wolf-Rayet bump regions (blue and purple inset windows, respectively). 
In the blue inset window, there are strong emission lines, but they appear narrow and nebular in
origin, whereas there is a lack of any features in the purple inset window for J141851 but a possible
\ion{C}{4} bump in J104457 which we discuss in \autoref{sec:whatbb}. }
\label{fig2}
\end{center}
\end{figure*}

MODS optical spectra of J104457 and J141851 were obtained on the UT dates of 2018 May 19 and 18, respectively.
We obtained simultaneous, moderate-resolution blue ($R\sim1850$) and red ($R\sim2300$) spectra.
J104457 and J141851 were each observed using the 1\arcsec\ longslit for 45 min.
Spectra were reduced with the MODS reduction pipeline\footnote{http://www.astronomy.ohio-state.edu/MODS/Software/modsIDL/}. 
One-dimensional spectra were corrected for atmospheric extinction and flux calibrated based 
on observations of flux standard stars \citep{oke90}. 
A portion of the resulting spectra are plotted in \autoref{fig2}, showing the broad spectra coverage provided
by this LBT/MODS setup, extending from $3200-10,000$ \AA\ and including a number of very high-ionization 
emission lines:
[\ion{Ne}{3}] \W3869, [\ion{Fe}{5}] \W4227, \ion{He}{2} \W4686, and [\ion{Ar}{4}] \W\W4711,4740.
While only the UV spectra can be used to determine the massive star populations, 
we can use the high-ionization UV and optical emission lines together to better 
constrain the shape of the ionizing spectral energy distribution.


\section{The Stellar Continuum}\label{sec:continuum}


\subsection{Fitting the UV Continuum}\label{sec:contmethods}

\begin{figure*}
\begin{center}
    \includegraphics[width=\textwidth]{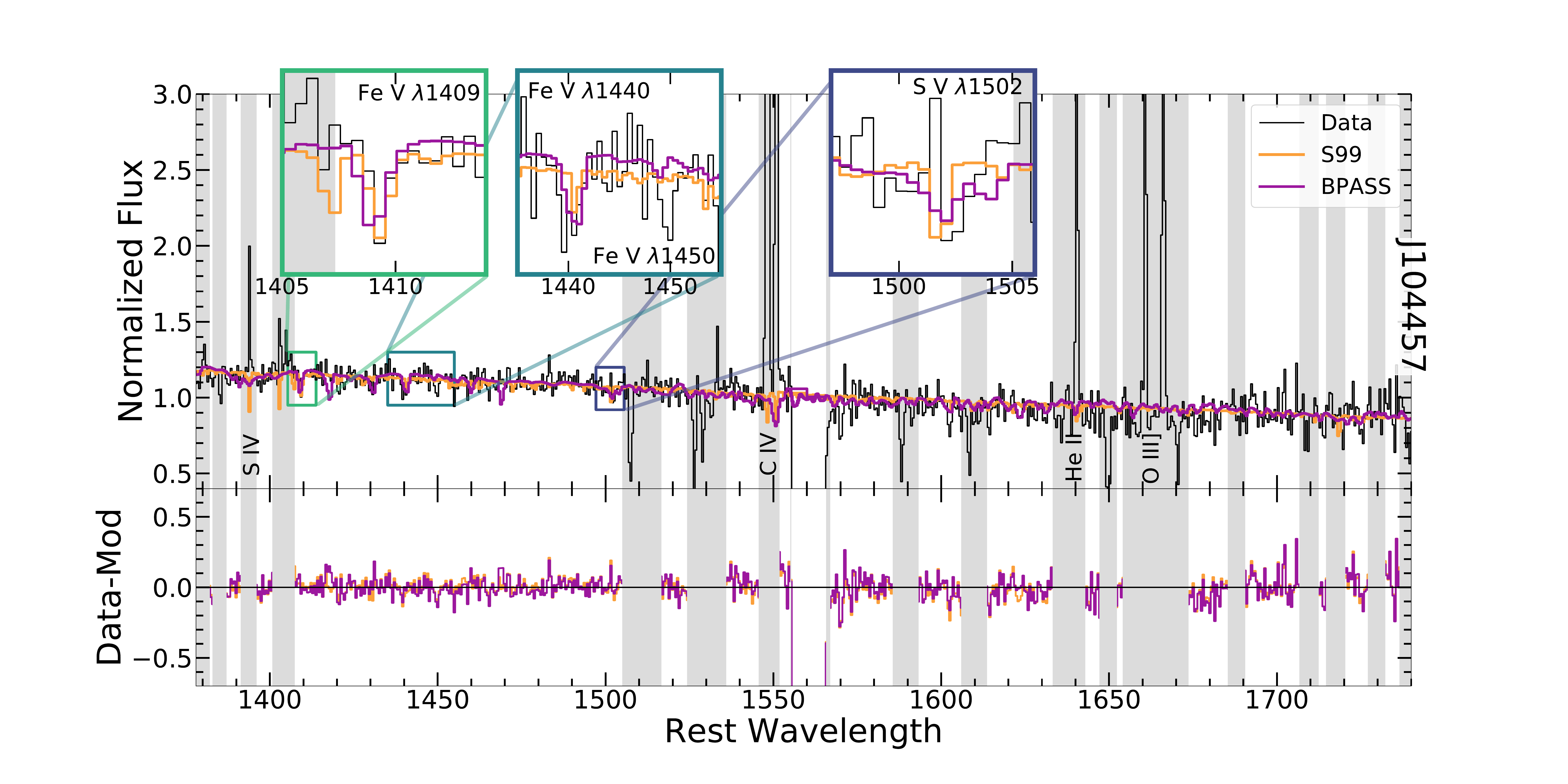}
    \includegraphics[width=\textwidth]{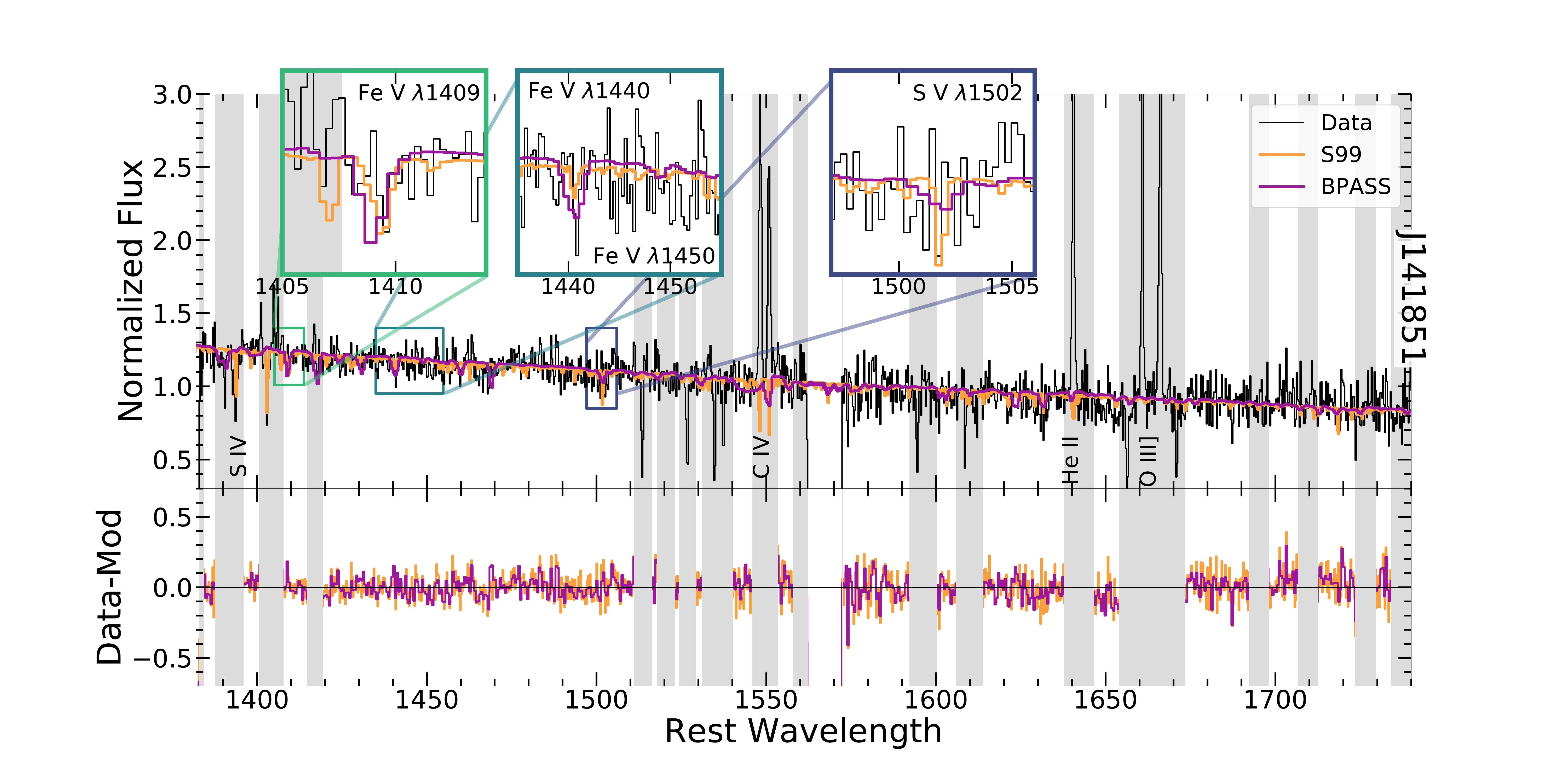}
\end{center}
\caption{The UV spectra of J104457 and J141851 and the best-fit continuum models. The data are plotted in black, and the spectra are normalized by their flux at 1495 \AA. We show the continuum fits from the Starburst 99 models \citep{leitherer99} in orange and BPASS models \citep{eldridge17} in pink for J104457 in the top plot and J141851 in the bottom plot. We also plot the residuals, continuum fit subtracted from the data, in the bottom panel of each plot. The zoom-in regions highlight detections of absorption lines from highly ionized metal absorption features -- \ion{Fe}{5} and \ion{S}{5} -- that are only present in the photospheres of the most massive and hottest stars. }  
\label{fig:J10spec}
\end{figure*}

We use the fitting procedure introduced in \cite{chisholm19} to determine the 
properties of the stellar populations in these galaxies.
We first mask out any features in the spectrum that are produced by non-stellar sources.
For example lines produced by Milky Way absorption and ISM emission and absorption 
lines associated with the galaxies are masked.
We used \cite{leitherer11} and \cite{bruhweiler81} as references to identify the lines to mask.
In addition to masking the spectral lines not associated with the stars, we masked the regions 
of the spectrum where the signal-to-noise was less than 1.
These masked regions are shown as the gray bands in \autoref{fig:J10spec}.
We also normalize the spectrum by the flux density at 1495 \AA\ in order to do our
fitting.

To obtain the age and metallicity of the stellar populations we compare the 
spectra to theoretical models of single-age bursts of star formation.
We test single star evolution using the high mass-loss rate stellar evolution models \citep{meynet94} from
Starburst99 (S99; \citealt{leitherer99}) and binary stars from BPASS v2.2.1 
\citep{eldridge17,stanway18} separately. 
We explore Kroupa initial mass functions \citep{kroupa02} with a broken 
power-law with a high and low-mass slope of 2.3 and 1.3, respectively, 
and both 100 M$_\odot$ and 300 M$_\odot$ high mass cut offs. 
Our model grids for both single stars and binary stars cover a range of 
metallicities from 0.05 Z$_\odot$ to 2 Z$_\odot$ (0.05, 0.2, 0.4, 1.0, 
and 2.0 Z$_\odot$) and ages from 1 Myr to 15 Myr in eight non-uniform 
steps (1, 2, 3, 4, 5, 8, 10, and 15 Myr) selected to capture variations 
in the FUV stellar continuum features with age.
We used these ranges of metallicities and ages because these galaxies 
appear to be young, metal-poor star forming objects based on their
extreme H$\beta$ equivalent widths and high star-formation rates 
\citep{berg16,berg19a}. 
We do not expect, nor do we find, that high metallicity or old stellar 
populations create the observed UV spectrum as it is dominated by the 
light from the most massive stars from young stellar populations, 
but we included the models to ensure we did not introduce this bias to our 
fitting method.

We convolve the 0.4 \AA\ resolution of the S99 single star models to the
spectral resolution of the individual galaxy spectra \citep{leitherer99,meynet94}.
For the binary stellar models from BPASS, the resolution is 1 \AA\ ,  so
our observed galaxy spectra were convolved to that larger resolution 
for these fits \citep{eldridge17}.
The fitting results are shown in \autoref{sec:spops}.

We fit the stellar continuum assuming that the FUV light is a linear 
combination of multiple single age and single metallicity bursts of star 
formation scaled by a linear coefficient. 
The final intrinsic stellar continuum fit is the sum of all 
models weighted by their coefficients.
We then fit for the best fit dust attenuation value (E(B-V)) by reddening 
the intrinsic spectrum using a foreground dust screen model 
and the \citet{reddy16} attenuation law. 
We tested using a \citet{calzetti00} and a SMC dust attenuation law \citep{gordon03},
but did not find that the continuum fits and stellar parameters, such as age and metallicity, statistically varied (although the E(B-V) parameter changed). 
The final fit has 41 free parameters (40 linear coefficients for the stellar 
continuum models and 1 dust attenuation model). 
We derive light-weighted stellar population parameters (i.e., age and metallicity) 
by calculating a weighted average using the model parameters and 
the best fit linear coefficients. 
In \autoref{tbl2}, we give the light-weighted parameters 
as well as the ionizing photon production efficiency ($\xi_{\rm ion}$; or the number of ionizing photons divided by the luminosity at 1500\AA). 

To find the errors on these parameters, we run the fitting code on
1000 randomized spectra.
During each randomized iteration, we created the randomized spectra by
multiplying the observed flux at a given wavelength by a random number with 
the distribution width dictated by the uncertainty on the spectrum at that wavelength.
For each iteration, we estimated the light-weighted properties and then took
the standard deviation of the distribution as the uncertainty on the stellar 
population properties.


\subsection{Stellar Continuum Parameters}\label{sec:spops}

The fits to the observed stellar continua provide estimates of the 
massive star population of these two EELGs. 
The first two rows in \autoref{tbl2} list the estimated light-weighted
ages and metallicities of these two galaxies.
We include four different sets of models that we used to fit the stellar 
populations: single-star evolution with a 100~M$_\odot$ upper IMF cutoff 
from S99 \citep{leitherer99,meynet94}, binary-star evolution 
with a 100~M$_\odot$ upper IMF cutoff from BPASS \citep{eldridge17}, 
single-star evolution with a 300~M$_\odot$ cutoff from S99
\citep{leitherer99,meynet94}, and binary-star evolution with a 300~M$_\odot$ 
upper IMF cutoff from BPASS \citep{eldridge17}. 
The fits imply that both galaxies have young (\simlt\ 10 Myr) and low-metallicity
($<0.15$~Z$_\odot$) stellar populations. 
Due to the modest signal-to-noise ratios, the inferred stellar population 
properties have relatively high errors, but the estimates using the four 
different model assumptions are consistent with each other within the 1$\sigma$
uncertainties. 
The BPASS models are not systematically older than the S99 models
for these two galaxies, likely because the fitting algorithm 
identified a feature corresponding to an older 
stellar population in J141851, pushing the S99 models to an older age.

As a measure of the total production of ionizing photons, we use the
stellar continuum fits to determine production efficiency of ionizing 
photons ($\xi_{\rm ion} = Q/L_{1500}$) by integrating the total number 
of ionizing photons from the fitted stellar population model ($Q$) from 
$1-912$ \AA\ and dividing by the FUV luminosity density at 1500 \AA.
We find log($\xi_{\rm ion}$) values of 25.82 and 25.68~dex [photons Hz erg$^{-1}$]
for J104457 and J141851, respectively, with relatively large errors (0.8~dex). 
These values are larger than (but statistically consistent with) the $\xi_{\rm ion}$ 
values inferred from the optical H$\beta$ flux ratios \citep{berg19a},
suggesting that the stellar population fits are producing an approximately 
sufficient number of hydrogen-ionizing photons to match the observed Balmer emission lines. 
The total ionizing photon production is consistent regardless of the 
assumptions made for the stellar models (single-star versus binary-star evolution
and the high-mass cutoff).
Since the S99 and BPASS models produce different parameters for their 
age and metallicity to recreate the observed spectrum,
it is surprising that $\xi_{\rm ion}$ is consistent between the two models (See \autoref{tbl2}).
However, the combination of older binary stars from BPASS produces a
similarly shaped ionizing continuum as the single-stars combined for S99,
thus the consistency of the $\xi_{\rm ion}$ values is less surprising.
Our high-mass cutoff models do produce $\sim$0.1 dex more ionizing photons
in our $\log \xi_{ion}$ calculations than our 100~M$_\odot$ models. 
This difference is similar to the expected difference between the 100 M$_\odot$
and 300 M$_\odot$ models from BPASS found in Figure 35 of \cite{eldridge17}.
However, the errors on our calculated numbers are 0.8 dex and thus the difference
in the 300~M$_\odot$ and 100~M$_\odot$ models is not significant.
The inclusion of the 300~M$_\odot$ only changes the fraction of single stars by 
a few percent in the BPASS models and does not imply a significant number of very
massive stars \citep{stanway19}.


\begin{deluxetable*}{c|cccc}
\tablecaption{Best-Fit Stellar Continuum Models}
\tablehead{
\CH{Model}  & \CH{S99 100 M$_\odot$} & \CH{BPASS 100 M$_\odot$} & \CH{S99 300 M$_\odot$}  &  \CH{BPASS 300 M$_\odot$}}
\startdata	
\bf J104457 \\
Age [Myr]                               & 1.04 $\pm$ 2.75 & 4.01 $\pm$ 1.13 & 1.26 $\pm$ 2.36 & 4.03 $\pm$ 0.85 \\
Metallicity [Z$_\odot$]                 & 0.09 $\pm$ 0.05 & 0.14 $\pm$ 0.04 & 0.08 $\pm$ 0.05 & 0.13 $\pm$ 0.03 \\
E(B-V)                                  & 0.25 $\pm$ 0.02 & 0.24 $\pm$ 0.02 & 0.21 $\pm$ 0.02 & 0.26 $\pm$ 0.02 \\
log($\xi_{ion}$) [phot Hz erg$^{-1}$]   & 25.9 $\pm$ 0.9  & 25.7 $\pm$ 0.8  & 25.9 $\pm$ 0.8  & 25.8 $\pm$ 0.8 \\
\hline
\bf J141851 \\
Age [Myr]                               & 10.59 $\pm$ 6.41 & 3.37 $\pm$ 1.24 & 9.69 $\pm$ 6.62 & 4.56 $\pm$ 1.08 \\
Metallicity [Z$_\odot$]                 &  0.06 $\pm$ 0.05 & 0.06 $\pm$ 0.04 & 0.08 $\pm$ 0.09 & 0.10 $\pm$ 0.03 \\
E(B-V)                                  &  0.16 $\pm$ 0.03 & 0.17 $\pm$ 0.03 & 0.13 $\pm$ 0.03 & 0.16 $\pm$ 0.03 \\
log($\xi_{ion}$) [phot Hz erg$^{-1}$]   &  25.7 $\pm$ 0.9  & 25.7 $\pm$ 0.8  & 25.7 $\pm$ 0.8  & 25.7 $\pm$ 0.8 \\
\enddata
\tablecomments{The stellar population parameters for the best-fit models to the 
UV continuum for each of the 4 types of models: 
S99 100M$_\odot$ - single star with 100 M$_\odot$ cut off for the IMF;
BPASS 100M$_\odot$ - binary star with 100 M$_\odot$ cut off for the IMF;
S99 300M$_\odot$ - single star with 300 M$_\odot$ cut off for the IMF;
and BPASS 300M$_\odot$ - binary star with 300 M$_\odot$ cut off for the IMF. 
Here we report the light-weighted ages and metallicities for both galaxies.
We used the \cite{reddy16} reddening law to determine the E(B-V).
We calculated the number of ionizing photons and report them here as log($\xi_{ion}$).}
\label{tbl2}
\end{deluxetable*}


\subsection{Characteristic Stellar Spectral Features}\label{sec:specfeat}

In addition to determining the stellar populations of these galaxies, 
the FUV observations of extreme emission line galaxies present a rare opportunity 
to study the massive star population properties at very low metallicity 
($<$10$\%$ Z$_\odot$). 
\autoref{fig:J10spec} reveals largely featureless
continua, with a very weak C~\iv~1550~\AA\ wind P-Cygni wind feature observed in 
J141851 and a nearly flat \ion{C}{4} region in J104457 as can be seen in the 
residuals in \autoref{fig:J10spec}. 
These weak winds are consistent with theoretical predictions of radiatively 
driven stellar winds, where mass-loss rates are significantly lower at extremely 
low metallicity \citep{vink}.  
While the continua are nearly featureless power-law spectra, at our 
signal-to-noise there are three identifiable weak photospheric features: 
\ion{Fe}{5}~1409+1440~\AA\ and \ion{S}{5}~1502~\AA\ 
\citep[see \autoref{table:3}; ][]{bruhweiler81, leitherer11}. 
The combination of the weak \ion{C}{4} P-Cygni features, small highly ionized
metal photospheric features, and an otherwise featureless stellar continuum are 
consistent with the young, low-metallicity stellar populations inferred from the
continuum fits.

\ion{Fe}{5}~1409+1440~\AA\ and \ion{S}{5}~1502~\AA\ absorption lines are 
indicative of very massive stars as only hottest stars keep
significant amounts of Fe and S in this high of ionization states.
If we observe absorption in \ion{Fe}{5} and \ion{S}{5}, this implies the most
massive stars are contributing most of the light for these galaxies, and the
massive stars only exist as the youngest stars.
In \autoref{table:3}, we compare the observed \ion{Fe}{5} and \ion{S}{5} equivalent
widths of our two EELGS to two samples of star-forming galaxies at $z\sim0$ 
\citep{heckman15, chisholm19} and $z\sim2$ \citep{rigby18a, rigby18b}, as well as the
stacked composites for both samples (bottom of each section). 
The \ion{S}{5} equivalent widths of both galaxies are significantly weaker than the 
comparison samples, likely because the metallicities of the EELGS are a factor of 5-10 
lower than the comparison samples. 
Interestingly, the \ion{Fe}{5}~1440~\AA\ detection in J104457, is among the strongest 
detections in the sample. 
While these two EELGS have low Fe-abundances \citet{berg21}, the \ion{Fe}{5} feature is actually 
stronger than the alpha-element tracer \ion{S}{5}.
This suggests that the iron opacity within the atmospheres of these low-metallicity 
stars is non-negligible. 
The only other spectra with \ion{Fe}{5}~1440~\AA\ strongly detected are J0150+1260, 
which is the second youngest galaxy in the \citet{chisholm19} low-redshift 
sample at 2.20 $\pm$ 2.17 Myr, 
and S000451.7-010321, which is dominated by a very young stellar population 
with a small contribution from an older population, 
suggesting that \ion{Fe}{5}~1440~\AA\ is a robust indicator of 
very young stellar populations even at low metallicity. 
This is consistent with the slightly older stellar population fit to J141851 and 
the non-detection of \ion{Fe}{5}~1440~\AA. 
These detections of Fe and S photospheric lines suggest that the UV dominant 
stars in these two EELGS are extremely young, massive, and low-metallicity. 


\begin{table*}[t]
\centering
    \begin{tabular}{cccc}
    \tablecaption{Photospheric Features}
    Object & \ion{Fe}{5} [\AA] & \ion{S}{5} [\AA] & Ratio \\
    \hline \hline
    J104457  & 0.24 $\pm$ 0.12 & 0.12 $\pm$ 0.11 & $\leq$ 0.52\\
    J141851  & $\leq$ 0.07 & $\leq$ 0.10 & $\leq$ 1.43 \\
    \hline \hline
    $z\sim0$\\
    J0021+0052  & $\leq$ 0.19 & 0.37 $\pm$ 0.21 & $\geq$ 1.95 \\
    J0150+1260  & 0.19 $\pm$ 0.02 & 0.35 $\pm$ 0.22 & 1.84 $\pm$ 1.17 \\
    J0808+3948  & N/A & 0.38 $\pm$ 0.09 & N/A \\
    J0824+2806  & $\leq$ 0.06 &  N/A & N/A \\
    J0926+4427  & $\leq$ 0.17 & $\leq$ 0.26 & N/A \\
    J0938+5428  & N/A & 0.30 $\pm$ 0.13 & N/A \\
    J1415+0540  & $\leq$ 0.15 & 0.47 $\pm$ 0.16 &  $\geq$ 3.13 \\
    J1416+1223  & $\leq$ 0.05 & N/A & N/A \\
    J1429+0643  & $\leq$ 0.21 & $\leq$ 0.22 & N/A \\
    J1429+1653  & N/A & $\leq$ 0.18 & N/A \\
    \hline 
    COS Stack   & $\leq$ 0.02 & 0.20 $\pm$ 0.03 & $\geq$ 10.00 \\
    \hline \hline
    $z\sim2$\\
    Cosmic Horseshoe        & $\leq$ 0.26 & 0.30 $\pm$ 0.08 & $\geq$ 1.15 \\
    RCS-0327-1326 Knot E    & $\leq$ 0.04 & 0.30 $\pm$ 0.06 & $\geq$ 7.50 \\
    S000451.7-010321        & 0.18 $\pm$ 0.04 & 0.16 $\pm$ 0.04 & 0.89 $\pm$ 0.30 \\
    S090003.3+223408        & $\leq$ 0.02 & 0.06 $\pm$ 0.03 & $\geq$ 3.00 \\
    Sunburst Arc Region 5   & $\leq$ 0.05 & 0.05 $\pm$ 0.02 & $\geq$ 1.00 \\
    \hline
    MegaSaura Stack         & $\leq$ 0.01 & 0.39 $\pm$ 0.02 & $\geq$ 39.00 \\
    \hline \hline
    \end{tabular}
    \caption{Table of equivalent widths of \ion{Fe}{5} 1440.528 \AA\ line and \ion{S}{5} 1501.76 \AA\ 
    line for our two new galaxies as well as two samples of star-forming galaxies at  
    $z\sim0$ \citep{heckman15, chisholm19} in the central section of the table 
    and $z\sim2$ \citep{rigby18a, rigby18b} in the bottom section.}
    \label{table:3}
\end{table*}


\section{Photoionization Modeling}\label{sec:photoion}


\subsection{Adopted Nebular Property Constraints}

Owing to the detection of a number of very high-ionization emission lines 
(e.g., \ion{C}{4}, \ion{He}{2}, \ion{O}{4}, \ion{Si}{4}, \ion{Ne}{5}, and \ion{Ar}{4}), 
we adopted a novel 4-zone ionization model in \citetalias{berg21} and recalculated the
nebular properties and chemical abundances of J104457 and J141851.
Therefore, for consistency, we adopt the results of \citetalias{berg21} as a better estimate 
of the nebular properties of J104457 and J141851 in our photoionization models. 
Here we briefly discuss the nebular properties that we use to constrain the shape 
of the ionizing spectral energy distributions of EELGs.

\subsubsection{Suite of UV$+$Optical Nebular Emission Lines}

In order to compare our observations to photoionization modeling, 
we first need to precisely measure a suite of UV and optical emission lines
that probe a large range in ionization potential energy.
We began by adopting the nebular emission line intensities measured in \citetalias{berg21}.
The detailed information on the line measurement and dereddening processes can be found in 
Section 2.3 of \citetalias{berg21}
and the subsequent emission line intensities in their corresponding Table 2.  
To briefly summarize, the UV and optical emission lines were measured in a uniform, consistent manner.
The optical emission lines were measured from continuum-subtracted LBT/MODS spectra 
using Gaussian profile fits.
The UV emission lines, on the other hand, were measured prior to continuum 
subtraction (as the UV continuum was not yet fit at the time of \citetalias{berg21})
and so used a local linear continuum fit and Gaussian emission profile.
Both UV and optical emission lines were corrected for Galactic extinction using
the \citet{green15} extinction map with a \citet{cardelli89} reddening law,
followed by an internal reddening correction determined from the Balmer decrement
and the \cite{cardelli89}/\cite{reddy16}) reddening law for the optical/UV lines.

The suite of emission line measurements we adopt for comparison with our 
photoionization modeling are reproduced in Table~\ref{table:suite}.
Specifically, we focus on the following emission lines for different ionization zones:
\begin{itemize} 
\item Low ionization zone: [\ion{O}{2}] \W\W7320,7330, [\ion{N}{2}] \W6584 
\item Intermediate ionization zone: \ion{C}{3}] \W\W1907,1909, [\ion{Ar}{3}] \W7136, [\ion{S}{3}] \W6312 
\item High ionization zone: \ion{O}{3}] \W\W1661,1666, [\ion{Ar}{4}] \W\W4711,4740
\item Very high ionization: \ion{O}{4} \W\W1401,1405,1407, \ion{C}{4} \W\W1548,1550, \ion{He}{2} \W1640, \W4686, [\ion{Ne}{3}] \W3869
\end{itemize}

\begin{table}[t]
\centering
    \begin{tabular}{c|c|cl}
    \tablecaption{Ionization Line Suite}
    Ionization Zone & Energy [eV] & Species & Wavelength [\AA\ ] \\
    \hline \hline
    Low             & $<$ 29 & [\ion{N}{2}]  & 6584 \\
                    &        & [\ion{O}{2}]  & 7320, 7330 \\
    \hline
    Intermediate    & 24-35  & [\ion{S}{3}]  & 6312 \\
                    &        & [\ion{Ar}{3}] & 7136 \\
    \hline
    High            & 35-55  & [\ion{O}{3}]   & 5007 \\
                    &        & [\ion{Ar}{4}] & 4740 \\
    \hline
    Very High       & $>$ 54 & \ion{He}{2}   & 1640 \\
                    &        & \ion{He}{2}   & 4686 \\
                    &        & [\ion{Ne}{3}] & 3869 \\
                    &        & \ion{O}{4}    & 1401, 1407 \\
    \hline \hline
    \end{tabular}
    \caption{The suite of ionization lines we use in this work
    to test the fit between the photoionizaton models and the 
    observations.
    }
    \label{table:suite}
\end{table}

\subsubsection{Ionization Structure}
In \citetalias{berg21}, we used the suite of nebular emission lines to measure
the ionization parameter $\log U$ for 
these two EELGs and found that the ionization structure is not 
well described by a single $\log U$ measurement from [\ion{O}{3}]/[\ion{O}{2}].
Classically, studies use 3 ionization zones to describe the ionization 
structure of a galaxy or an \ion{H}{2} region.  
Using the 4-zone ionization model, we were able to
constrain the $\log U$ for three ionization zones of the nebula
to better understand the high-, intermediate-, and low-ionization 
energy photons separately.
As a result, we adopted these three ionization parameters from 
Table 3 of \citetalias{berg21} for comparisons with
our photoionization modeling.


\subsection{Photoionization Modeling} \label{sec:photmods}

The richness and quality of our combined UV and optical spectra of J104457
and J141851 allowed us to perform the first detailed models of these spectra 
with the goal of simultaneously constraining the direct-method gas-phase 
properties, inferred ionizing stellar populations, and the resulting nebular 
emission lines they produce.
To do so, we used \texttt{CLOUDY} version C17.01 \citep{ferland17} to run 
models with radiation field inputs from 
(1) single-aged bursts of binary star formation from the BPASS 2.2 models \citep{eldridge17,stanway18} in \S \ref{sec:mod1}, 
(2) the stellar population continuum fits from Section \ref{sec:continuum} in \S \ref{sec:mod2}, and
(3) combination models of the fitted stellar populations plus an extra high energy source modeled 
as a blackbody in \S \ref{sec:mod3} in order to recreate our very high energy ionization lines.


\subsubsection{Single-Aged Burst Binary Models}\label{sec:mod1}

For our fiducial model grid, we used BPASS binary, single-aged stellar 
population models with ages ranging from 1-10 Myr and spanning a metallicity range 
of $0.005Z_\odot < Z < 0.3Z_\odot$ \citep{eldridge17,stanway18}.
We matched the gas-phase metallicity to the stellar metallicity in the models,
as expected for alpha-elements, but investigated sub-solar abundance patterns of 
carbon and nitrogen relative to oxygen following the empirical trends seen for metal-poor 
dwarf galaxies \citep[e.g.,][]{vanzee06,berg12,berg19b}.
\citetalias{berg21} investigated the possibility of enhanced $\alpha$/Fe 
ratios in our EELGs but found no evidence that the inclusion of increased
$\alpha$-elements were capable of producing the high-energy photons that 
are missing from these galaxies.
Specifically, we varied the fraction of C/O and N/O relative to solar from 0 to 1 in 
steps of 0.25, encompassing the measured abundances of N/O = 0.44 N/O$_\odot$ and 
0.32 N/O$_\odot$ and C/O = 0.42 C/O$_\odot$ and 0.35 C/O$_\odot$ for J104457 and 
J141851, respectively.
Finally, we used a variety of constant densities ranging from 10 cm$^{-3}$ to 
10,000 cm$^{-3}$ in logarithmic steps. 
We also used power law density gradients of the form $n(r) = n_0(r_0) (r/r_0)^\alpha$
where $n_0$ is the density at the illuminated surface of the cloud 
which we set to 10 cm$^{-3}$, 100 cm$^{-3}$, 1000 cm$^{-3}$, and 10,000 cm$^{-3}$, 
and where $r_0$ is the radius from the source to the illuminated surface of the cloud,
and slopes of $\alpha=$[-1,-2,-3].

We compare our models to the observed spectra of J104457 and J141851 by comparing 
the emission line ratios of different species.  
For our study we use eight emission lines from the optical spectrum relative to
H$\beta$: [\ion{O}{2}] \W\W7320,7330, [\ion{N}{2}] \W6584, [\ion{Ar}{3}] \W7136, 
[\ion{S}{3}] \W6312, \ion{O}{3}] \W\W1661,1666, [\ion{Ar}{4}] \W\W4711,4740, 
[\ion{Ne}{3}] \W3869, and \ion{He}{2} \W4686; and two UV lines relative to 
[\ion{O}{3}] \W1666: \ion{He}{2} \W1640 and [\ion{O}{4}] \W\W1401,1407.
We use these lines from the optical spectrum to cover emission lines produced
by ionization zones from low to very-high ionization and are not highly temperature
and density dependent, except for [\ion{O}{2}] which we use because we also use
\ion{O}{3}] and [\ion{O}{4}].
We only use the \ion{He}{2} and [\ion{O}{4}] lines from the UV spectrum because
the only other available lines are \ion{C}{3}] and \ion{C}{4} which we do not use
because of inconsistencies in the C/O abundance measurement.

\begin{figure*}
\begin{center}
    \includegraphics[width=1.0\textwidth]{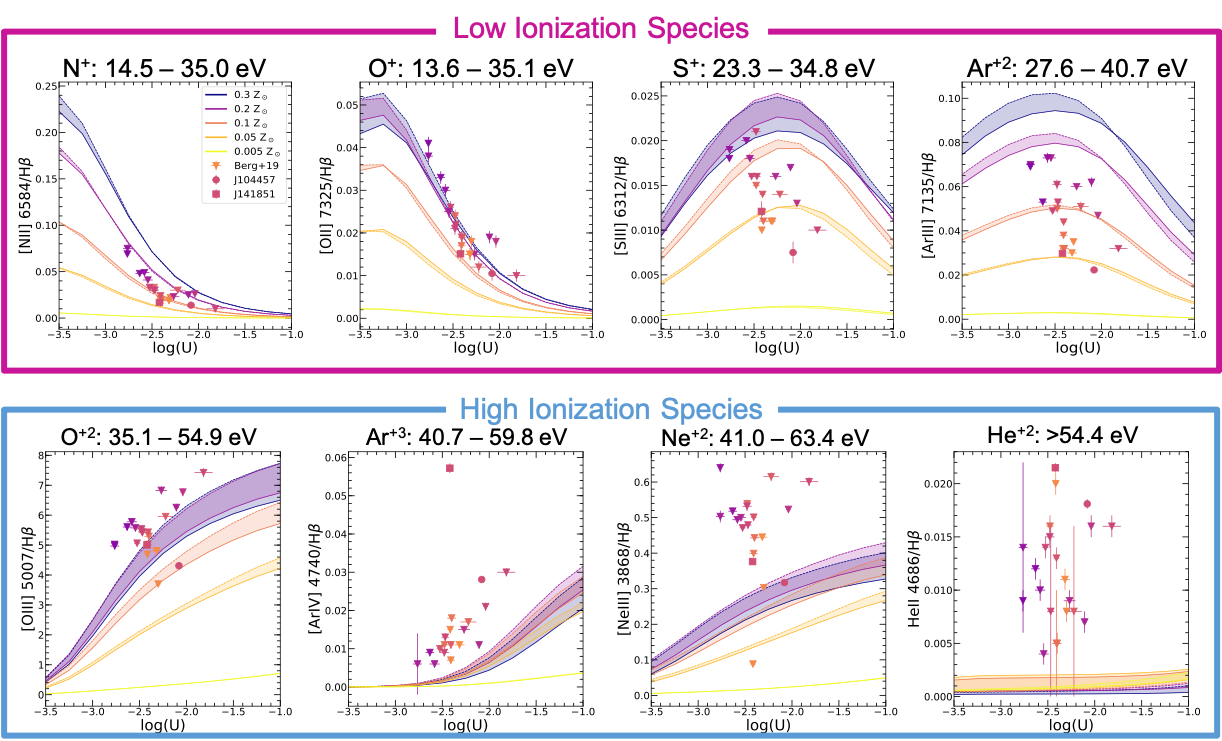}
    \caption{
    \texttt{CLOUDY} photoionization models of several nebular emission line ratios resulting from BPASS single-age stellar populations. 
    The models are color-coded by their metallicity and are plotted as shaded bands to represent the range of ages,
    with the lower edge at 1 Myr and the upper edge at 3 Myr.  
    For comparison, we plot the observed line ratios from J104457 (circle point), J141851 (square point),  
    and the \cite{berg19b} sample (triangles). 
    The points are color coded according to the measured nebular metallicity of each galaxy, along the same color scale as the models.
    The line ratios plotted are separated by their ionization potential energies such that low-ionization species are in the top row and
    high-ionization species are in the bottom row, demonstrating the progressive discrepancy between
    models and observations with higher ionization potential.}
    \label{fig:bpassvobs}
\end{center}
\end{figure*}

Similar to \cite{berg18} and others, we start by first finding the best 
fit model from our fiducial grid by using a reduced-$\chi^2$, $\chi^2_{\rm red}$ fit. 
We compare the line ratios of low ionization energy species from J104457 and 
J141851 and the sample of small star forming galaxies presented in \cite{berg19b} 
to these single age bursts and find good agreement with the models.
Specifically, our models fit [\ion{N}{2}], [\ion{O}{2}], [\ion{S}{3}], 
and [\ion{Ar}{3}] well.
In the top row of \autoref{fig:bpassvobs} we have plotted the BPASS 
single age burst models from 1 Myr (solid lines) to 3 Myr (dashed lines) for 
metallicities ranging from 0.005 - 0.3 Z$_\odot$. 
The points for each galaxy are color coded by their metallicity.
The results of the reduced $\chi^2$ of each fit ($\chi^2_{\rm red} =
\chi^2/$d.o.f.), the logU, the metallicity, the age, the C/O, and the N/O ratios
are listed in the section labeled 4.2.1 in \autoref{table:J10fixZfits}
and \autoref{table:J14fixZfits} for J104457 and J141851 respectively.
Although these models fit the low- and intermediate-ionization lines 
relatively well, the lines with ionization energies through [\ion{Ar}{3}], 
the $\chi^2_{\rm red}$ is very large for the high- and very high-ionization lines, 
lines with ionization energies for [\ion{O}{3}] and above ($>$ 35 eV). 
The single-burst models do not match the high-ionization emission lines. 

When we compare the high and very-high ionization energy lines ratios, 
\ion{O}{3}], [\ion{Ar}{4}], [\ion{Ne}{3}], \ion{He}{2} $\lambda$4686,
\ion{He}{2} $\lambda$1640, and [\ion{O}{4}], predicted by the single-age 
burst models to the observed ratios from J104457, J141851, and the 
\cite{berg19b} sample, we find an offset in the line ratios.  
The observed high-ionization lines are systematically higher than the 
ratios predicted by the single-age bursts. 
We attempted changing a number of additional parameters in our models
including ionization parameter, density, and power-law density 
distributions in order to recreate the observations, but these changes
were still not enough to reproduce the observed line ratios.
We show this offset in the lower row of \autoref{fig:bpassvobs}.
We conclude that single-age bursts of star formation as predicted by 
BPASS do not produce enough high energy photons to recreate the observed
highly ionized gas.


\begin{table*}[t]
\centering
    \begin{tabular}{c|c|cccccccc}
    \tablecaption{J104457 Fit Parameters}
    Model & Ionization Zone & \multicolumn{8}{c}{J104457 = 0.058 Z$_\odot$} \\
    \hline \hline
    Section & and (lines used) & $\chi^2$ & logU & Z (Z$_\odot$) & Age (Myr) & $\frac{C/O}{(C/O)_\odot}$ & $\frac{N/O}{(N/O)_\odot}$ & Temp (kK) & BB \% \\
    \hline
    \ref{sec:mod1}      & Low (O2,N2)                & 0.29    & -2.5  & 0.05 & 5.01 & 0.25 & 0.25 & - & -  \\
    Single-Aged Burst   & Inter. (S3,Ar3)            & 0.71    & -3.0  & 0.05 & 10.0 & 1.25 & 1.25 & - & -  \\
    Binary Models       & High (O3,Ar4)              & 42.36   & -1.0  & 0.05 & 1.0  & 0.25 & 0.25 & - & -  \\
                        & V. High (Ne3,He2,He2,O4)   & 294.46  & -1.0  & 0.15 & 3.16 & 1.25 & 1.25 & - & -  \\
    & {\bf All 10}      & {\bf 151.14} & {\bf -1.0} & {\bf 0.15} & {\bf 3.16} & {\bf 1.25} & {\bf 1.25} & {\bf -} & {\bf -} \\
    \hline
    \ref{sec:mod2}      & Low (O2,N2)                & 2.03    & -2.5 & 0.05  & {\it 4.01} & {\it 0.34} & {\it 0.32} & - & - \\
    Stellar Population  & Inter. (S3,Ar3)            & 1.14    & -3.25 & 0.05  & {\it 4.01} & {\it 0.34} & {\it 0.32} & - & - \\
    Fits Models         & High (O3,Ar4)              & 2.68    & -1.75 & 0.07  & {\it 4.01} & {\it 0.34} & {\it 0.32} & - & - \\
                        & V. High (Ne3,He2,He2,O4)   & 363.51  & -1.0  & 0.058 & {\it 4.01} & {\it 0.34} & {\it 0.32} & - & - \\
    & {\bf All 10}      & {\bf 174.10} & {\bf -1.75} & {\bf 0.07} & {\bf 4.01} & {\bf 0.34} & {\bf 0.32} & {\bf -} & {\bf -} \\
     \hline
    \ref{sec:mod3}          & Low (O2,N2)                & 1.90    & -2.25 & {\it 0.058} & {\it 4.01} & {\it 0.34} & {\it 0.32} & 60  & 80 \\
    Combination Stellar $+$ & Inter. (S3,Ar3)            & 2.94    & -3.5  & {\it 0.058} & {\it 4.01} & {\it 0.34} & {\it 0.32} & 60  & 0  \\
    Blackbody Models        & High (O3,Ar4)              & 9.68   & -1.75 & {\it 0.058} & {\it 4.01} & {\it 0.34} & {\it 0.32} & 80  & 10 \\
                            & V. High (Ne3,He2,He2,O4)   & 1.71    & -1.25 & {\it 0.058} & {\it 4.01} & {\it 0.34} & {\it 0.32} & 80  & 50 \\
    & {\bf All 10}  & {\bf 30.40} & {\bf -1.75} & {\bf 0.058} & {\bf 4.01} & {\bf 0.34} & {\bf 0.32} & {\bf 80} & {\bf 60} \\
    \hline
    \end{tabular}
    \caption{
    J104457 parameters of best fit stellar population and gas from models described in \autoref{sec:photmods}. All parameter values in \textit{italics} denote that the parameter value was fixed for the model, additionally the age, (C/O), and (N/O) ratios were fixed for the All 10 line model for the 4.2.2 models, and the metallicity, age, (C/O), and (N/O) ratios were fixed for the All 10 line model for the 4.2.3 models. Photoionization models from \autoref{sec:mod1} use single age bursts of star formation from BPASS v2.1 as the input spectrum with gas metallicites fixed to the same as the stars, but variable (C/O) and (N/O) ratios.  Models from \autoref{sec:mod2} use the ionizing continuum from \autoref{sec:spops} as the input spectrum with gas phase metallicites varying around the gas phase metallicity measured in \citetalias{berg21} and (C/O) and (N/O) ratios fixed to the abundances from \citetalias{berg21}.  The models from \autoref{sec:mod3} use the ionizing continuum from \autoref{sec:spops} plus an ad hoc blackbody as the input spectrum with fixed gas phase metallicity and (C/O) and (N/O) ratios as measured with the 4-zone abundances in \citetalias{berg21}.}
    \label{table:J10fixZfits}
\end{table*}

\begin{table*}[t]
\centering
    \begin{tabular}{c|c|cccccccc}
    \tablecaption{J141851 Fit Parameters}
    Model & Ionization Zone & \multicolumn{8}{c}{J141851 = 0.087 Z$_\odot$}\\
    \hline \hline
    Section & and (lines used) & $\chi^2$ & logU & Z (Z$_\odot$) & Age (Myr) & $\frac{C/O}{(C/O)_\odot}$ & $\frac{N/O}{(N/O)_\odot}$  & Temp (kK) & BB \%  \\
    \hline
    \ref{sec:mod1}      & Low (O2,N2)                 & 17.54  & -2.25  & 0.1  & 3.16 & 0.25 & 0.25 & -  & -      \\
    Single-Aged Burst   & Inter. (S3,Ar3)             & 0.01   & -1.25  & 0.15 & 5.01 & 0.5  & 0.5  & -  & -      \\
    Binary Models       & High (O3,Ar4)               & 338.50 & -1.0   & 0.1  & 1.0 & 1.0 & 1.0 & -  & -      \\
                        & V. High (Ne3,He2,He2,O4)    & 465.14 & -1.0   & 0.15 & 3.16 & 0.25 & 0.25 & -  & -      \\
    & {\bf All 10}      & {\bf 315.99} & {\bf -1.0} & {\bf 0.15} & {\bf 3.16} & {\bf 1.25} & {\bf 1.25}& {\bf -} & {\bf -} \\
    \hline
    \ref{sec:mod2}      & Low (O2,N2)               & 51.97   & -2.25 & 0.07  & {\it 3.37} & {\it 0.31} & {\it 0.40} & - & -       \\
    Stellar Population  & Inter. (S3,Ar3)           & 1.18    & -2.5  & 0.05  & {\it 3.37} & {\it 0.31} & {\it 0.40} & - & -       \\
    Fits Models         & High (O3,Ar4)             & 9.11    & -1.25 & 0.07  & {\it 3.37} & {\it 0.31} & {\it 0.40} & - & -       \\
                        & V. High (Ne3,He2,He2,O4)  & 558.75  & -1.0  & 0.087 & {\it 3.37} & {\it 0.31} & {\it 0.40} & - & -       \\
    & {\bf All 10}      & {\bf 308.38} & {\bf -1.75} & {\bf 0.11} & {\bf 3.37} & {\bf 0.31} & {\bf 0.40} & {\bf -} & {\bf -} \\
     \hline
    \ref{sec:mod3}          & Low (O2,N2)                 & 68.98  & -2.25 & {\it 0.087} & {\it 3.37} & {\it 0.31} & {\it 0.40} & 80  & 80       \\
    Combination Stellar $+$ & Inter. (S3,Ar3)             & 3.81   & -1.75 & {\it 0.087} & {\it 3.37} & {\it 0.31} & {\it 0.40} & 100 & 10       \\
    Blackbody Models        & High (O3,Ar4)               & 18.75  & -1.75 & {\it 0.087} & {\it 3.37} & {\it 0.31} & {\it 0.40} & 100 & 70       \\
                            & V. High (Ne3,He2,He2,O4)    & 3.19   & -1.5  & {\it 0.087} & {\it 3.37} & {\it 0.31} & {\it 0.40} & 80  & 70       \\
    & {\bf All 10}      &{\bf 62.03} & {\bf -1.75} & {\bf 0.087} & {\bf 3.37} & {\bf 0.31} & {\bf 0.40} & {\bf 80} & {\bf 70} \\
    \hline
    \end{tabular}
    \caption{
    Same as \autoref{table:J10fixZfits} but for J141851. All parameter values in \textit{italics} denote that the parameter value was fixed for the model, additionally the age, (C/O), and (N/O) ratios were fixed for the All 10 line model for the 4.2.2 models, and the metallicity, age, (C/O), and (N/O) ratios were fixed for the All 10 line model for the 4.2.3 models.}
    \label{table:J14fixZfits}
\end{table*}


\subsubsection{Stellar Population Fits Models}\label{sec:mod2}

The single-age bursts of star formation underpredict the high energy 
ionization lines as shown in \autoref{sec:mod1}. 
We have more knowledge of the stellar populations in J104457 and J141851 
than most EELGs because we have observations of the FUV stellar continuum 
which we fit in \autoref{sec:continuum}.  
As a result, we can test whether the observed stars are capable of producing
the high-energy ionization lines by using the ionizing spectrum from the 
stellar populations discussed in \autoref{sec:spops} as the input for our
\texttt{CLOUDY} models.  
We fix our stellar parameters to those produced in \autoref{sec:spops} for
these models but we vary the gas phase parameters to match the nebular 
properties of 104557 and J141851 published in \citetalias{berg21}. 
Just as was done for the fiducial models, both constant density and 
power-law density distributions were considered. 
The affects of a power-law density distribution are negligible for
reproducing high ionization emission and thus we only present the results
from our constant density models.

\begin{figure*}
\begin{center}
    \includegraphics{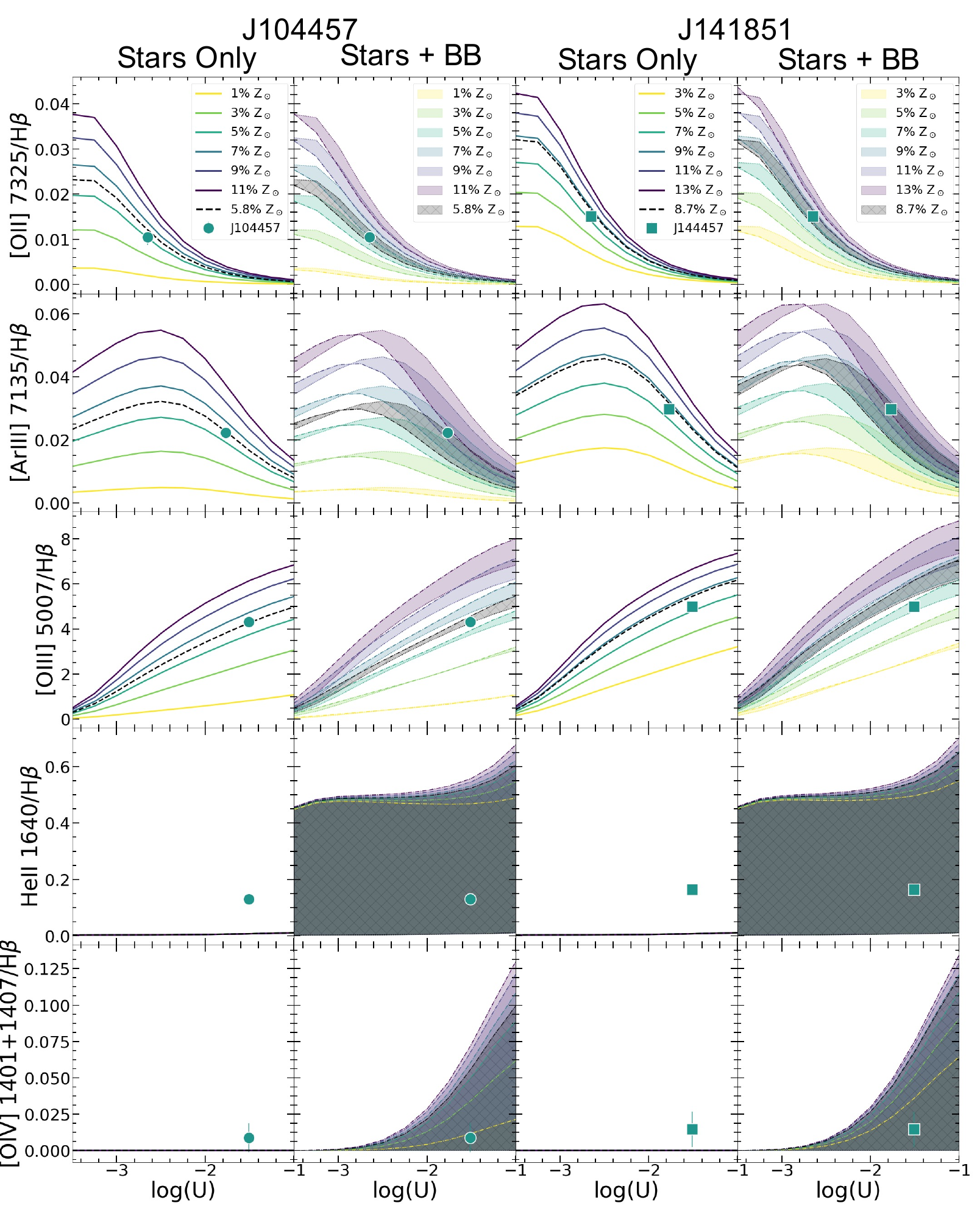}
    \caption{Line ratio diagrams for 5 emission lines spanning the different ionization zones with the observed line ratio plotted as the point for J104457 and J141851. The "Stars Only" column shows the \texttt{CLOUDY} photoionization models where we used the ionizing continuum from the stellar population we fit to these galaxies in \autoref{sec:spops} as the input ionizing spectrum. Each line is a different metallicity from low metallicity at 1\% Z$_\odot$ in yellow to 11\% Z$_\odot$ in purple for J104457 and for J141851 the range starts with 3\% Z$_\odot$ in yellow and goes to 13\% Z$_\odot$ in purple. The black dashed line shows the results of a model run with the observed gas-phase metallicity of each galaxy, 5.8\% Z$_\odot$ for J104457 and 8.7\% Z$_\odot$ for J141851. In the "Stars + BB" column the models are plotted in bands with the same metallicity color-coding system, but the bands now include the effects of the blackbodies we modeled. The band is bordered by a dotted line showing the  0\% $L_{\rm young}$ (no BB) model and a dashed-dotted line showing the 80\% $L_{\rm young}$ from a 100kK blackbody. The black hatched region shows the models run with the observed gas-phase metallicity of the galaxy.}
    \label{fig:mega}
\end{center}
\end{figure*}

We also specialized our fits for J104457 and J141851 by adjusting the C/O and
N/O ratios to the values reported in \citetalias{berg21} to see if these models would
reproduce the high-ionization lines we observed.
We set the metallicity of the gas to the metallicity measured in \citetalias{berg21}
and then vary the metallicity with linear steps of 0.02 (2\% $Z_\odot$).
The results of these models can be seen in the first (J104457) and third (J141851) 
columns of \autoref{fig:mega}. 
We plot the metallicity steps around the observed metallicity as colorful lines
and the observed metallicity as the black line for each galaxy.  

As we did with the models in \autoref{sec:mod1}, we fit the models to the 
observed line ratios from the 8 optical lines shown in \autoref{fig:bpassvobs} 
and 2 UV lines.  
These results are shown in the Section 4.2.2 region of \autoref{table:J10fixZfits} and 
\autoref{table:J14fixZfits}.  
Similar to the models from the previous section, the models that are specialized 
using the observed stellar population and gas phase abundances do a reasonable job
of reproducing the low- and intermediate-ionization lines, but fail to reproduce 
the very-high ionization lines.
The stellar populations we observed are able to match the high-ionization lines 
\ion{O}{3}] and [\ion{Ar}{4}] that the single-age bursts could not match well, 
as seen by the lower $\chi^2_{\rm red}$ values for the high ionization zone for 
Section 4.2.2 in comparison to Section 4.2.1 in \autoref{table:J10fixZfits} and 
\autoref{table:J14fixZfits}.
Although the stellar populations from \autoref{sec:spops} can recreate the {low-}, 
intermediate-, and high-ionization lines, these models are unable to produce the 
observed values for the very high-ionization energy lines like \ion{He}{2} and 
[\ion{O}{4}].


\subsubsection{Combination Stellar + Blackbody Models}\label{sec:mod3}

The stellar population fit models from the previous section 
failed to reproduce the very high-ionization emission lines as seen in the 
\ion{He}{2} and [\ion{O}{4}] panels of "Stars Only" columns of \autoref{fig:mega}.  
This problem is often seen for \ion{He}{2} observations 
(e.g. \citealt{kehrig15,kehrig18,kehrig21,senchyna17,senchyna20,wofford21}), 
but we emphasize that the discrepancy between photoionization models using the stellar 
continuum as the input ionizing spectra and observations of emission lines is a 
problem for all very-high ionization lines, especially \ion{He}{2} and [\ion{O}{4}], 
as seen in \S\ref{sec:mod1} and \S\ref{sec:mod2} above. 
Here we explore what is required to reconcile the observed nebular emission lines
with the models.
In the absence of new, specialized models of low metallicity stellar populations,
we can provide a first bench mark to improve these models using the addition of
a blackbody (BB).

We combine the models from \autoref{sec:mod2} with a hot BB in order 
to reproduce the unexplained excess flux we observe in the very high-ionization lines.
We added BBs with temperatures of 60,000 K, 80,000 K, and 100,000 K because
these temperatures span the range between the energies dominated by X-ray emission
and the energies observed in the UV.  
\cite{senchyna20} demonstrated that similar objects do not have significant 
X-ray emission so we avoid adding high temperature BBs or a power-law contribution
that would be detected in X-rays.
 
We added these three BBs (T=60kK, 80kK, 100kK) in varying strengths by 
including BBs with 10\% steps of the total fractional luminosity produced by the 
stellar population fit in Section \ref{sec:continuum}. 
We refer to this quantity as the luminosity of the young stars ($L_{\rm young}$). 
We add this BB to the input spectrum of the young stars and then vary the 
log$U$ of the photoionization models using the combined spectra as the shape of
the SED for the ionizing source. 

We ran photoionization models with the same gas metallicity steps as those in 
\autoref{sec:mod2}; this includes the observed gas-phase metallicity from 
\citetalias{berg21} and the 2\%$Z_\odot$ steps around that measured metallicity.  
We used the abundances from the 4-zone ionization structure as estimated 
by \citetalias{berg21} to fix our C/O and N/O ratios.  

We plot the effects of the BB on each metallicity model in the columns
labeled "Stars + BB" of \autoref{fig:mega}.
The lower bound (dashed line) of each metallicity shows the 60kK 10\% BB
while the upper bound (dotted line) plots the 100kK 80\% BB so each band
covers the entire range produced by our different BB possibilities. 
The black band on this plot shows the models run with the metallicity fixed 
at the observed metallicity.  

We see that the added BB increases the range of line ratios each 
metallicity model is capable of producing. 
For the observed metallicity (black band), we see that similar to the models
from the previous two sections, we are still able to reproduce the emission 
from the low-, intermediate-, and high-ionization lines, which demonstrates 
that the added BB does not greatly affect the lower energy parts of 
the ionizing spectrum.  
Additionally, the very high-ionization lines now are included within the band
the models produce.  
This shows us that the additional BB produces enough ionizing photons 
at energies $>$54 eV to reproduce the observed \ion{He}{2} emission.

\begin{figure*}
\begin{center}
    \includegraphics[width=\textwidth]{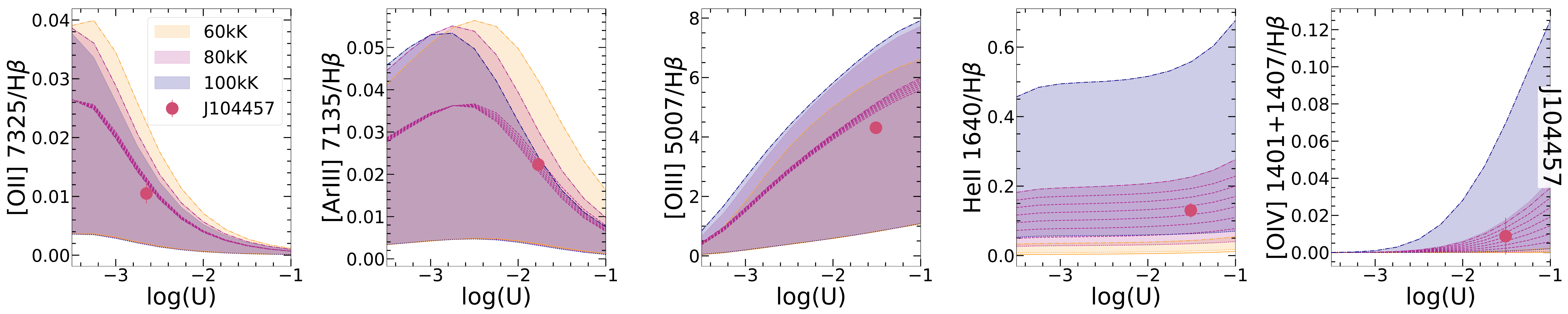}
    \includegraphics[width=\textwidth]{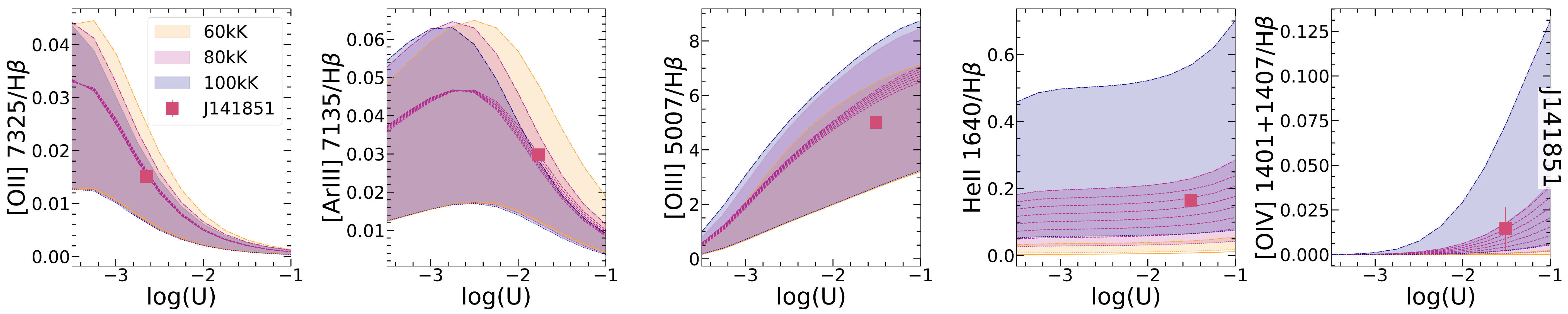}
    \caption{Combined stellar population and BB models for J104457 and J141851 to show the effects of BB temperature and percentage. Each temperature is plotted as a different color band with 60kK as orange, 80kK as pink, and 100kK as blue. The bands range from the lower dotted line as a BB with 0\% $L_{\rm young}$ to upper dashed-dotted line showing 80\% $L_{\rm young}$ and include all metallicity steps from the models.  The pink dashed lines show the 10\% $L_{\rm young}$ steps from 10$-$80\% for an 80kK blackbody run with the observed gas phase metallicity for each galaxy. The point is the observed line ratio for the galaxy.}
    \label{fig:bbhoriz}
\end{center}
\end{figure*}

The shape of the BB is determined by the temperature and percent 
contribution to the luminosity. 
We show the effects of the BB temperature on the models in 
\autoref{fig:bbhoriz}, each temperature is a different color and the shaded
region spans from a lower bound of 10\% BB to an upper bound of 80\% BB.  
We plot the observed line ratios for J194457 in the top panels and J141851 in the bottom panels.
The low- and intermediate-ionization lines have the ability to give us more 
information about the temperature of the BB.
High ionization parameters with large line ratios can rule out certain BB temperatures.
The high-ionization lines provide more information about the percent contribution.
The shapes of the BB models allow for any temperature to produce
the very-high emission lines.
At high line ratios, higher temperature
BBs can produce the same emission with a smaller percent contribution.  
To demonstrate this for the \ion{He}{2}/H$\beta$ plot we show dashed lines 
of each additional 10\% in the percentage contribution to luminosity for 
the 100kK BB.
This shows that with larger percent contributions the BB
can reach increasingly extreme line ratios.

The very-high ionization zone includes the most information about the BB
component of the fit as the \ion{He}{2} and [\ion{O}{4}] lines are most sensitive
to the high energy photons.  
J104457 requires an 80kK BB with 60\% of the young stellar light while J141851
requires an 80kK BB producing 70\% of the young light.  
We can see for J104457 in \autoref{fig:bbhoriz} all three BB 
temperatures are capable of reproducing the observed line ratios for our low- 
and intermediate-ionization lines.
For J141851 in \autoref{fig:bbhoriz}, we see that the [\ion{Ar}{3}] line starts to fall outside the
band of line ratios produced by the 100kK BB, indicating a 100kK
BB may not produce the emission we see from J141851.
In J104457, the low and intermediate zone fits do not contain much 
information that constrains the BB components, but in J141851 these
lines may contain some temperature information.

Our solutions employ a number of different selections of the 8 optical lines and 
2 UV lines we've been using in this paper.  
The fit results for J104457 and J141851 are shown in 
\autoref{table:J10fixZfits} and \autoref{table:J14fixZfits} respectively.  
We report the temperature of the BB and the percentage of total 
light that the BB makes up in this table in addition to the columns
discussed in \autoref{sec:mod1} and \autoref{sec:mod2}.  
The $\chi^2_{\rm red}$ values for the combined fit from all 10 emission
lines are high because the combined fits do not fit all 10 lines well
simultaneously, particularly [\ion{O}{3}] increases our $\chi^2_{\rm red}$
values.
The combined fit does favor the BB parameters fit using only the 
very-high ionization lines which gives us confidence the BB 
information from all 10 lines is accurate, since the BB information
matches the parameters found by the lines with the most information about the
BB.  
When we compare the fits that include this BB to the fits from
\autoref{sec:mod1} and \autoref{sec:mod2}, we see that the added BB allows
for a better fit when using lines from the very-high ionization zone and in the 
total 10 line fit of all ionization zones.
This demonstrates that including a high energy photon source strongly improves 
the predicted nebular emission structure of very-high ionization emission line galaxies.


\section{Discussion} \label{sec:discuss}


\subsection{The High-Energy Ionizing Photon Production Problem}

J104457 and J141851 are two EELGs that demonstrate the inability of traditional photoionization
models to accurately reproduce the very high-energy ionization lines.  
We show this discrepancy in \autoref{sec:photmods} with the very high-ionization 
lines of \ion{He}{2} and [\ion{O}{4}].  
In addition to the excess \ion{He}{2} and [\ion{O}{4}] emission we see compared to 
the photoionization models without an additional source of high energy photons, 
J104457 and J141851 show this discrepancy in their [\ion{Fe}{5}] emission.  
\citetalias{berg21} detailed the ionization correction factors (ICFs) for these two galaxies 
and found that the observed [\ion{Fe}{5}] emission was not matched by the ICF provided 
from photoionization modeling with only the observed stars producing the photons.  
This inability of the stellar populations to reproduce the high-energy
photons through photoionization indicates there is a \textit{high-energy ionizing 
photon production problem} (HEIP$^3$) as named by \citetalias{berg21}.

The HEIP$^3$ is observed as excess \ion{He}{2} emission in these two EELGs
but also in the extreme \ion{He}{2} emission observed in other blue compact dwarf galaxies 
(e.g., \citealt{kehrig15,kehrig18,kehrig21,berg18,senchyna17,senchyna20,stanway20}).
\citetalias{berg21} showed that the HEIP$^3$ results in underestimated ionization parameters
which, when corrected, could increase the \ion{He}{2} emission produced by photoionization 
models used to compare with the extreme \ion{He}{2} emission in blue compact dwarf galaxies.
We demonstrated that even this increased ionization parameter could not account for the
\ion{He}{2} emission we observe in our two EELGs when only the stars are ionizing the gas,
so it is unlikely that correcting the ionization parameter would reconcile the excess \ion{He}{2} 
emission observed in other samples of blue compact dwarf galaxies or EELGs.
We have found that the addition of a BB to the observationally-motivated model
of the stars can solve the HEIP$^3$ for J104457 and J141851.  
This additional source of very-high energy photons can reproduce the very-high ionization
lines that are not reproduced as a result of the HEIP$^3$ as is described in 
\autoref{sec:selfcon}.  


\subsection{Self-Consistent Modelling Success} \label{sec:selfcon}

The emission lines from EELGs trace the ionizing spectrum of these galaxies, and, by using
photoionization modeling, we are able to compare different forms of stellar population
synthesis to the observed emission lines in order to understand what is ionizing these galaxies. 
We tested a number of input spectra to use in our \texttt{CLOUDY} photoionization modeling to reproduce 
the observed emission lines from these two galaxies. 
Using only the canonical models of single-age bursts of star formation, we were able to recreate 
the emission lines from low-energy ionization species, but we were unable to produce the 
high-energy ionization lines with these input as seen in \autoref{fig:bpassvobs}. 
\autoref{fig:bpassvobs} compares emission lines from not only J104457 and J141851, 
but also a number of compact dwarf galaxies in the nearby universe. 
Notably, the models of high-ionization lines fall short of the observations from all 
galaxies we plotted.
This indicates that models with a single-age burst of star formation do not produce the correct
ionizing continuum shape required to reproduce the observations of EELGs.  

We then tested the observed stellar populations, as we measured them from 
Section \ref{sec:continuum}, as the input spectrum. 
This moved us from using sets of likely stellar populations to the stellar 
population that we observed in each galaxy. 
These stellar populations were determined using models with and without binaries, 
as well as with an extended high-mass cutoff.  
The shapes of the ionizing continuum from the single-star and binary models were 
similar, so we used the binary model version as it has better wavelength resolution 
over the wavelength range we used to calculate the ionizing spectrum. 
With this observationally-motivated ionizing spectrum, we were able to reproduce the 
observed emission lines up to ionization energies of $\sim$35 eV, like OIII], which 
can be seen in the "Stars Only" columns of \autoref{fig:mega}. 
The stellar model fits to the observed stars in these galaxies produce enough photons 
to recreate the high-energy emission lines but not the very high-energy emission lines 
like \ion{He}{2} and [\ion{O}{4}].  
This inability to reproduce the highest energy emission lines demonstrates the HEIP$^3$ 
as introduced in \citetalias{berg21}. 

Since the models fit to the observed stars did not reproduce the highest energy emission
lines, we decided to add a BB to the input spectrum in order to get an idea of the 
temperature and percent contribution of the source needed to recreate these lines.  
The ad hoc BB has a peak with a temperature less than what is observable in the 
X-rays because \cite{senchyna20} and others show that EELGs like our sample
do not tend to have significant X-ray detections.  
The BB temperature must also be high enough to not disturb the low-energy 
ionization lines that are already produced by the observed stellar population.  
With these restrictions, we fit using 8 optical lines from all 4 ionization zones and 
found that the best fit temperature of the BB was 80,000 K and the flux 
contribution for the BB was 60-70\% of the total luminosity of the young stars, 
$L_{\rm young}$.
Full details of the fits can be seen in \autoref{table:J10fixZfits} 
and \autoref{table:J14fixZfits}.  
In order to understand more about our BB, we fit the models to 
multiple smaller sets of emission lines.
The low- and intermediate-ionization lines carry little information about the 
BB but do favor ionization parameters indicative of their ionization zones.  
In \autoref{fig:bbhoriz}, we see that the high- and very high-ionization lines hold 
most of the information about the BB and thus determine the BB 
parameters for the full 8 line fit of all ionization zones. 
This added BB successfully reproduces the low- and intermediate-ionization 
lines already produced with observed stars as well as the very-high energy lines 
that were missed.  

\begin{figure}
\begin{center}
    \includegraphics[width=\columnwidth]{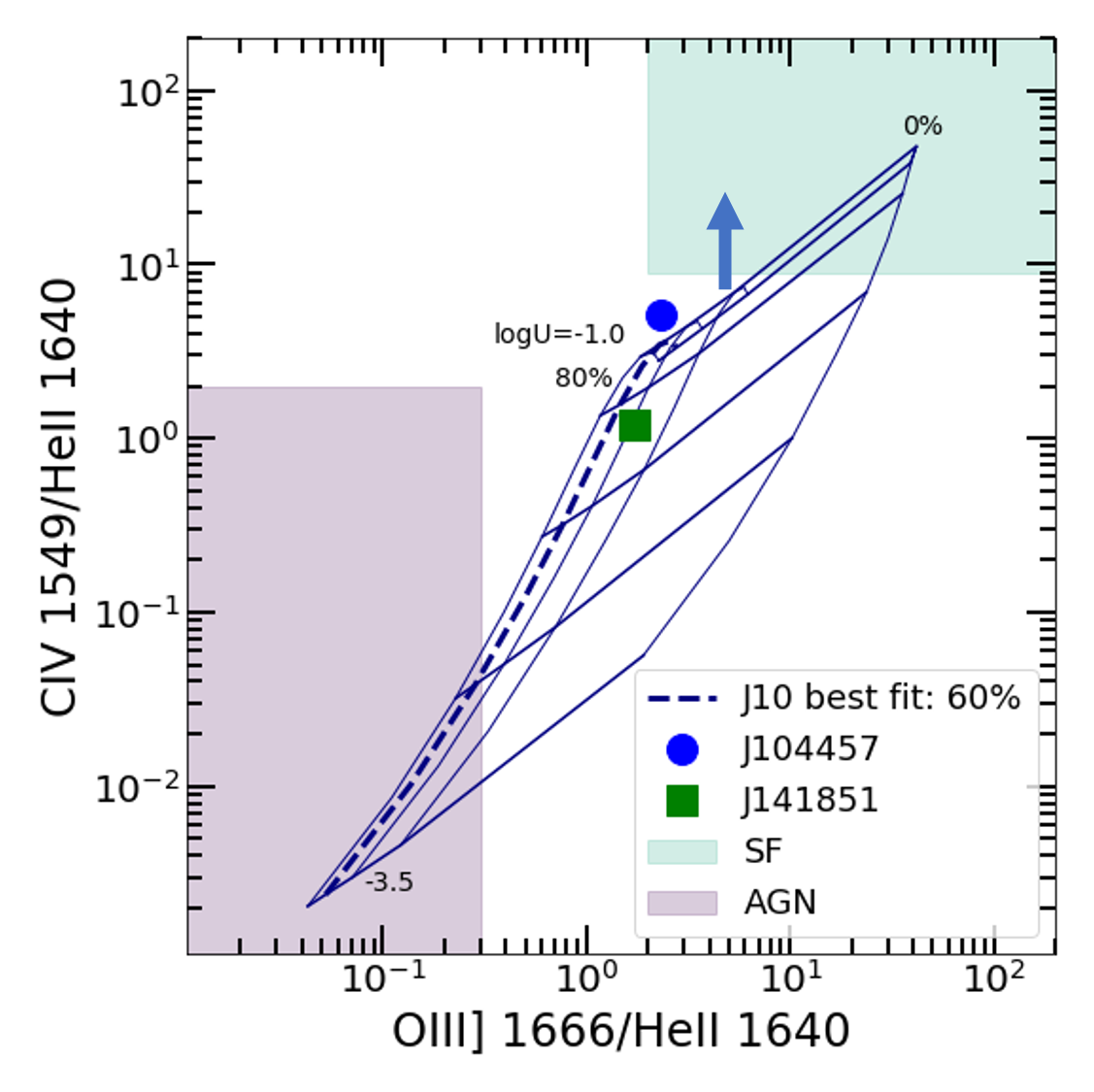}
    \caption{The UV BPT diagram plotted with \ion{C}{4} \W\W1548,50/\ion{He}{2} \W1640
    versus \ion{O}{3}] \W1666/\ion{He}{2} \W1640.
    The teal and purple shaded boxes highlight the regions typical of photoionization models
    for star-formation only and AGN only, respectively.
    In comparison, we plot the models for J104457, where the right-most line maps
    the models produced by the stellar population fits for a range of ionization parameters,
    extending from log$U=-1$ at the top to log$U=-3.5$ at the bottom with horizontal 
    lines marking steps of 0.5 in log$U$.
    Models adding contributions from an 80kK blackbody are plotted as narrow lines,
    where each subsequent model to the left increases the contribution from the blackbody by 20\%. 
    The arrow indicates the direction the models should move to account for the discrepancy between
    abundances calculated from ICFs using continuous and burst star-formation models.}
    \label{fig:uvbpt}
\end{center}
\end{figure}

The addition of a BB to understand the the ionizing radiation needed to 
produce the very-high ionization lines we observe in our EELGs is not a new idea. 
\cite{steidel14} used a similar BB to model the ionizing radiation of 
star-forming galaxies at $z\sim2-3$.
We compare our method to previous models by plotting our models of the observed
stellar population and ad hoc BB at fixed metallicity of 0.058 $Z_\odot$ 
for J104457 as a grid on a UV diagnostic diagram as seen in Figure \ref{fig:uvbpt}.  
The UV BPT diagram has been suggested as a way to determine if a galaxy is 
star-forming or AGN driven for galaxies at higher redshift where the UV diagnostic 
lines are the only observed diagnostics \citep{feltre16}.  
We use a diagnostic diagram consisting of \ion{C}{4} \W\W1548,50/\ion{He}{2} \W1640
versus \ion{O}{3}] \W1666/\ion{He}{2} \W1640 to distinguish the star-forming
and AGN driven objects.
The general range of star-formation models produced in \cite{mainali17} is shown 
as the teal shaded box on the upper right corner of the plot, while the AGN 
driven models are shown in the shaded purple box on the lower left \citep{feltre16}.   
Our models are plotted as a grid where the observed stellar population only is the 
rightmost line and each vertical line to the left is a 20\% jump in relative flux 
contribution of the BB. 
The horizontal lines are the log$U$ of the model which start at log$U$=$-$1.0 at 
the top and each horizontal line is a step of 0.5  in log$U$ to the bottom-most 
line of log$U$=$-$3.5. 
These models move across the gap from star-forming into the AGN driven regions 
of parameter space which covers the observations of these galaxies well.  

The observation for J104457 lies above the model grid in \autoref{fig:uvbpt}.  
This discrepancy between the models and the observations stems from the method 
for calculating the abundances based on the ionization correction factor (ICF).
We used abundances calculated by \citetalias{berg21} who used continuous star-formation
models to determine the ICF for carbon in our galaxies.
Continuous star-formation models under-predict the ICF when compared to 
models with bursts of star-formation, which leads to lower carbon abundances.
With increased C/O values, our models would increase along the 
\ion{C}{4} \W\W1548,50/\ion{He}{2} \W1640 axis and thus encompass the 
observation of J104457, as indicated by the arrow in \autoref{fig:uvbpt}.
In similar diagnostic diagrams using \ion{C}{4} \W\W1548,50/\ion{He}{2} \W1640
versus \ion{C}{3}] \W1666/\ion{He}{2} \W1640 we see the same effects;
our models do not produce enough C emission and by increasing the C/O abundance
our best-fit model would match the observations for these galaxies.

Our combined stellar population and ad hoc BB models span the
gap between AGN and star-formation driven models.
The space between the classical AGN and star-formation
models is primarily where we find EELGs like J104457 and J141851.
Since local EELGs lie between AGN and star-formation models,
observations of high-redshift galaxies likely fall in this
parameter space as well.
We caution that reality for high-redshift galaxies will not
conform to only star-formation or AGN models and going forward
a composite spectrum similar to the ones developed here should
be used.

In summary, the models of the observed stellar continuum are combined with an 
additional hot BB component in order to simultaneously fit the emission 
lines from all 4 ionization zones.  
This combination model reproduces the emission of \ion{He}{2} and [\ion{O}{4}] 
from the very high-ionization zone as well as [\ion{O}{2}] and [\ion{N}{2}] 
from the low-ionization zone and lines from the zones in between.
This observationally-motivated stellar continuum model and ad hoc BB 
are capable of reproducing the emission lines from all 10 emission lines
used in our fit.


\begin{figure*}
\begin{center}
    \includegraphics[width=\textwidth]{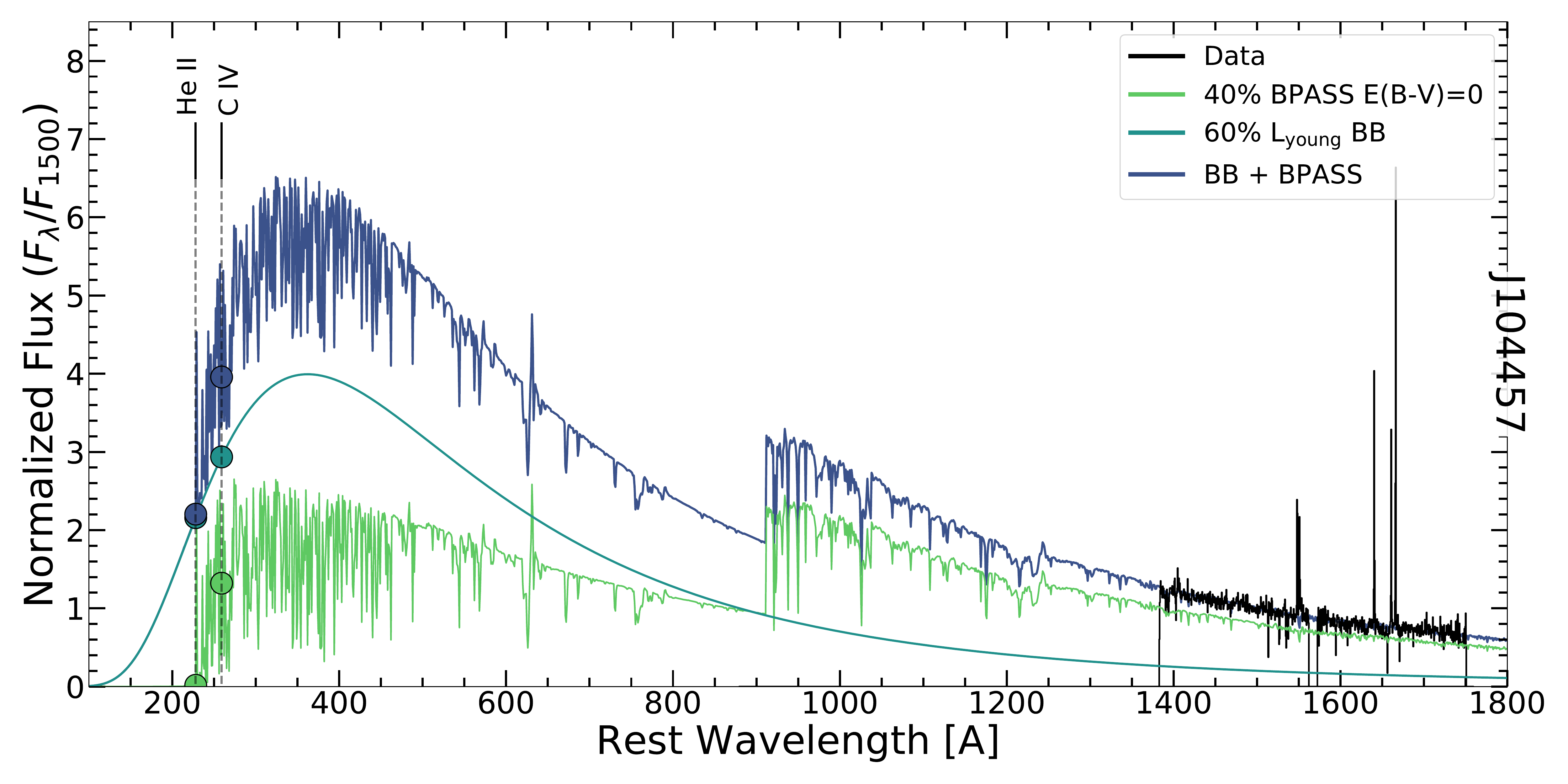}
    \includegraphics[width=\textwidth]{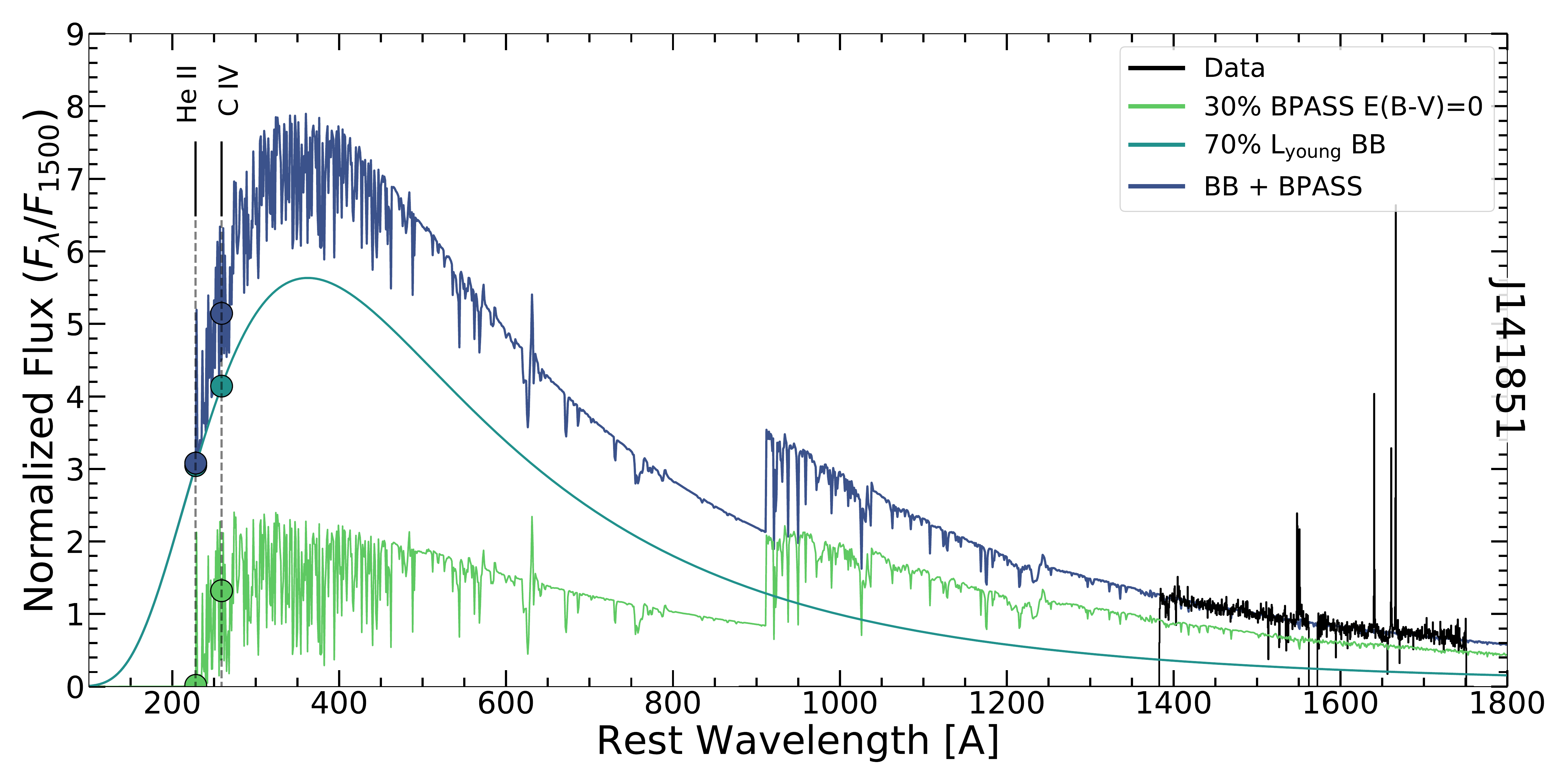}
    \caption{Ionizing continuum from the stellar continuum fits for the BPASS models for J104457 in the top panel and J141851 in the bottom panel. Includes a blackbody with a luminosity 60\% of $L_{\rm young}$ for J104457 and 70\% of $L_{\rm young}$ for J141851, a temperature of 80kK and the stellar continuum fit plus the plotted blackbody. The circles show the intersection of each model with the wavelength required to produce HeII and CIV.}
    \label{fig:j10ion}
\end{center}
\end{figure*}


\subsection{What is the source of the high energy photons?} \label{sec:whatbb}

Our BB and observationally-motivated stellar continuum 
combined model successfully recreates the high energy line ratios 
observed in our two EELGs. 
The ionizing continuum of the stars, BB, and combined 
stars and BB are shown in \autoref{fig:j10ion}. 
The green line shows the ionizing continuum from the best fit 
stellar population using BPASS models as described in Section
\ref{sec:continuum} scaled to account for BB contribution. 
The black spectrum shows the observed data, while the teal line shows the 
BB, and the blue line shows the stars and BB combined. 
The circles show the intersection of each model with the wavelength
required to produce \ion{He}{2} and \ion{C}{4} emission and are 
color-coded by model.

The BBs shown in \autoref{fig:j10ion} contribute a significant portion of the flux required 
to reproduce the high energy ionization lines and create a distinctly 
different shape in the far UV for these galaxies.
The sources(s) of the high energy photons are, to date, unidentified.  
We have explored the literature and found a number of candidates 
that could cause the very-high ionization emission lines we observe, and 
we discuss them and their feasibility below. 

\textit{Binaries.} 
In \autoref{sec:spops}, we tested binaries in our stellar population 
calculation by using the BPASS models and found that binary star models
recreate the observed UV continuum with similar precision but at 
slightly older ages. 
Even with the binary population models we cannot reproduce the BB 
emission; however binary models at lower metallicity provide 
more uncertainty in their models as a result of the 
unknown binary fractions as a function of metallicity, initial 
periods, and mass ratios which are measured from the local universe
but not in low metallicity environments \cite{stanway20}.

\textit{Low-metallicity stellar atmospheres.} 
These galaxies are very metal poor. 
As reported in \cite{berg16} for J104457 and \cite{berg19a} for J141851, 
the gas in these galaxies is only 5.8\% and 7.1\% Z$_\odot$ respectively 
and the stars themselves only 9-14\% and 6\% Z$_\odot$ as reported in Table \ref{tbl2}.
The stellar population synthesis models for low metallicity stars are
theoretical or extrapolated to lower metallicities since there are not 
observations of individual stars with $Z < 0.1 Z_\odot$
\citep{lejeune97,grafener02,smith03,lanz03,topping15}.  
However, stellar atmospheres are complicated structures that non-linearly 
depend on the temperatures, densities, mass-loss rates, and chemical abundances.
Slight variations in these parameters could produce dramatically different 
amounts of ionizing photons. 
Theoretical work on Population III stars indicates that they could 
provide more high-energy photons if they experienced significant mass-loss
\citep{schaerer02}. 
Observations of massive stars with $Z < 0.1 Z_\odot$ will provide insight 
as to whether low metallicity stars are capable of producing the emission 
we model as a BB in this work.  
Telford et al. (2021, accepted) shows that very low-metallicity
stars are not capable of producing enough high-energy photons to reproduce
extreme emission lines like \ion{He}{2}.

\textit{Very massive stars.} 
Another option to produce our BB radiation is very massive stars, stars 
that are $>$100$M_\odot$, which could be more plausible in the low 
metallicity regime.  
Very massive stars have been detected in SBS0335-052E by \cite{wofford21} 
who claim the models of very massive stars reproduce the UV and optical 
emission lines they have observed. 
Additionally, \cite{senchyna20} used a sample of 10 star-forming 
galaxies from the Sloan Digital Sky Survey to conclude that the 
discrepancy between stellar population models and the observed UV
lines could only be rectified with additional very massive stars.
When we tested both S99 and BPASS models with stellar populations up to 
300$M_\odot$ our ionizing continua did not sufficiently change.
The ionizing continuum was larger for the 300$M_\odot$ models, but
was not sufficient in reproducing the ionizing photons required in our BB.

\textit{Stripped stars.}
Another hot candidate for the source of the high energy photons is stripped stars 
in binaries as suggested by \cite{gotberg19,gotberg20}. 
These stripped cores of stars are extremely hot and produce significant \ion{He}{2} ionizing 
photons, but they require an older stellar population.  
As shown in \autoref{tbl2} the stellar populations found using the
UV spectra are only 1 - 10 Myr old.
This does not preclude an 
older stellar population existing in these galaxies, but the young stellar
population is clearly dominant in the emission lines we observe. 
\textit{WISE} photometry of J104457 indicates the infrared SED is dominated
by dust heated by the starburst and does not clearly show an older stellar
population \citep{brammer12}.

\textit{Wolf-Rayet Stars.}
Another metallicity dependent candidate are Wolf-Rayet (WR) stars. 
WR stars drive their envelopes off of the stellar core using line-driven 
winds, but at low metallicities there are few metals to absorb the momentum 
required to accelerate the wind.  
In addition, we do not observe any WR bumps in our spectrum for J141851, 
but there is a possible \ion{C}{4} bump that could signify WRs present 
in J104457, as evidenced in the inset panels of \autoref{fig2} as the
emission lines from most of those species appear narrow and nebular 
in origin.  
The existence of some WRs in J104457 could help partially solve our
missing high-energy photon problem, but with only one WR emission line
signature, it is unlikely we have a significant enough population to
fully solve our problem.
Finally, even when WR stars are observed in metal poor galaxies, 
like IZw18 and SBS0335-052E, they do not produce enough ionizing photons 
at energies high enough to produce \ion{He}{2} emission as shown in 
\cite{kehrig15,kehrig18}.
With all of these things considered, WRs are not likely the source of
our missing high-energy photons.

\textit{Active Galactic Nuclei (AGN).}
Since these high energy emission lines mimic those produced by AGN,
a reasonable candidate for our BB is a low luminosity AGN. 
However, we ran a photoionization model with an AGN spectrum and found
that this spectrum still overproduces the very-high ionization 
lines when scaled so that the low-, intermediate-, and high-ionization lines
match the observations, implying the spectrum is too hard 
to reproduce our suite of emission lines. 
The hardness of the AGN spectrum also implies the existence of 
hot gas that could be detected in the X-rays, but these dwarf galaxies
are rarely observed with X-ray detections \citep{senchyna17}.
The stellar masses of these galaxies are only 10$^{6.5-7}$ M$_\odot$ 
which according to relations of the host mass to central black hole
mass like the one produced in \cite{reines15} would imply an 
intermediate mass black hole to power the AGN.  
Intermediate mass black holes are notoriously difficult to observe;
however X-ray surveys including the \textit{Chandra} COSMOS-Legacy 
survey have allowed studies to find AGN in dwarf galaxies out to 
$z \leq$ 2.4 \citep{mezcua18}. 
Similar radio surveys have observed compact radio sources consistent
with AGN activity in dwarf galaxies \citep{reines20}, and through 
resolved emission line studies from the optical \citep{moran14,mezcua20}. 
\cite{mezcua18} suggest that the AGN fraction decreases to $\sim$ 0.4 
percent for dwarf galaxies with stellar masses in the range of 
$10^9 < M_* \leq 3 \times 10^9$ M$_\odot$ with respect to both X-ray
luminosity and stellar mass, which makes our $\sim 10^{6.7}$ M$_\odot$
galaxies unlikely to have AGN driving their emission lines.

\textit{Ultra-Luminous X-ray Sources (ULXs).}
Black holes could still cause the \ion{He}{2} emission we see if 
they are in the form of high-mass X-ray binaries or ULXs where 
a black hole or neutron star is in a binary with an O or B 
type star and accretes material from the stellar wind of the companion.
These accreting compact objects could produce the photons necessary
to excite \ion{He}{2} as suggested in \cite{schaerer19}; however 
the spectral shape of high-mass X-ray binaries is debated.
\cite{senchyna20} finds the ULX spectrum to be too hard 
for most small metal-poor galaxies as these galaxies do not tend to 
show X-ray detections in the high-energy bands which are usually 
produced by ULXs.
Recently, however, \cite{simmonds21} performed a similar analysis 
using ULX spectra based on observed ULXs in nearby galaxies and found
that the softer spectra are capable of reproducing \ion{He}{2} emission,
but they include the caveat that ULX spectra as a whole are not 
well understood.
In particular, \cite{schaerer19} proposed the \ion{He}{2} emission 
from IZw18 was produced by X-ray binaries, however, when \cite{kehrig21} 
explored the feasibility of this claim, they found a separation
of $\sim$200 pc between the high-mass X-ray binary and the \ion{He}{2} 
emission, again suggesting the X-ray binary is not responsible for the
\ion{He}{2} emission observed.

\textit{Supersoft X-ray Sources (SSSs).}
As ULXs are too hard, perhaps the younger sibling of compact 
object accretion, supersoft X-ray sources, where a white dwarf accretes 
material, could be the physical mechanism behind the high energy photons we require. 
SSSs have temperatures of $\sim$10-100 eV and bolometric luminosities 
of $\sim 10^{36}-10^{38}$ erg s$^{-1}$ and so, could possibly provide the 
very-high energy photons we need \citep{greiner91,kahabka94}. 
SSSs have not been observed in similar galaxies to-date; however the 
spectrum of these accreting WDs should fall into a similar parameter space 
as our BB. SSSs are rarely observed in nebulae similar to those found in 
J104457 and J141851 \citep{woods16}, and, as they are often viewed as the
progenitors for Type Ia SNe, it is important to note that Type Ias 
are rarely found in low-metallicity environments \citep{kobayashi98}.  
Additionally, the presence of WDs implies an older stellar population 
than we observe in the UV, and as mentioned with the stripped stars, 
the IR information from \textit{WISE} does not provide enough information
to conclude whether there is an older stellar population residing in these galaxies. 
Even with an older stellar population we would need $\sim$50,000 and 
$\sim$10,000 SSSs for J104457 and J141851 respectively to produce 
the luminosity of our BB (J104457: $5.1 \times 10^{42}\mathrm{erg s}^{-1}$; 
J141851: $1.1 \times 10^{42} \mathrm{erg s}^{-1}$). 
The spatial distribution of these SSSs would be random in the location
where the older stellar population formed, so a spatial analysis of the
\ion{He}{2} emission could help determine if SSSs provide the very-high 
energy photons we require.

\textit{Shocks.}
Shocks are another suggestion to produce the high-energy photons we need.  
Similar galaxies, IZw18 and SBS0335-052E, have soft X-ray emission that 
could correspond to shocks, but when the soft X-ray emission is integrated,
it does not account for enough energy to produce the \ion{He}{2} observed 
in those galaxies \citep{kehrig15,kehrig18,kehrig21}.  
Shocks should also produce emission lines from species like [\ion{Ne}{5}]
up to a line ratio of [\ion{Ne}{5}] \W3426/H$\beta$ < 0.3 for a 
shock propagating through a medium with density N$_e$ = 1 cm$^{-3}$ 
\citep{izotov12}.
We have detections for the [\ion{Ne}{5}] \W3426 line from both 
J104457 and J141851 and the [\ion{Ne}{5}] \W3426/H$\beta$ ratio 
for J104457 [\ion{Ne}{5}] \W3426/H$\beta$ is only 0.0011 and for 
J141851 [\ion{Ne}{5}] \W3426/H$\beta$ = 0.0060, these are ratios
are much smaller than the predicted models for shocks from \cite{izotov12}. 
Additionally, \cite{berg18} tested models with pure shock emission to 
explain the extreme \ion{He}{2} emission from the lensed star-forming 
galaxy SL2S J021737-051329 at $z \sim 2$ and found that shocks 
generally produced too little \ion{C}{3}] emission in these EELGs.

\textit{Catastrophic Cooling.}
Finally, suppressed super winds causing catastrophic cooling could 
cause very-high energy ionization line emission as suggested by 
\cite{gray19} and \cite{danehkar21}.
In this case, the rapid gas cooling through lines like \ion{O}{6} 
creates extreme emission beyond the capabilities of the stellar
population.  
\cite{gray19} predict significant \ion{He}{2} emission at the 
outer boundary of their nebula and \cite{danehkar21} predict 
\ion{O}{6} emission from the central parts of the nebula.  
Future work with spatially resolved \ion{He}{2} emission could help 
determine if these confined winds are the source of the additional 
high-energy photons we observe. 
Additionally, future observations of these galaxies should look for
emission lines indicative of this catastrophic cooling such as 
\ion{Si}{3} \W1206 and \ion{O}{6} \W1037.
We examined existing low resolution spectra for \ion{O}{6} for 
J104457 and J141851 and find that \ion{O}{6} is not 
significantly detected.

This is an incomplete list of possible candidates that could generate the high energy 
photons we infer. 
Future work can help determine which mechanism provides the
very-high energy photons. 
Spatial resolution of the \ion{He}{2} emission and 
observations of the X-rays produced by J104457 and J141851 will help rule out
some of the options provided in this list.
We hope that by providing a BB with temperature and percentage 
of the young stellar population L$_{\rm young}$ we are laying down a foundation for 
building a physical model of the source of the high-energy photons.


\section{Conclusions} \label{sec:conclusions}

We used the \textit{HST}/COS and LBT/MODS spectra presented in 
\citetalias{berg21} to study the stellar populations and 
extreme emission lines from two \textit{extreme emission line galaxies}
(EELGs), J104457 and J141851.
These EELGs have the largest reported \ion{C}{4} and \ion{He}{2}
EWs in the local universe which suggests these are likely
reionization analogue galaxies.
We fit the stellar populations for these galaxies using
the UV continuum from the \textit{HST}/COS spectrum
and find these are very young stars with ages $\leq$ 10 Myr
and metal-poor with metallicities $\leq$0.15 Z$_\odot$.
We used photoionization modeling to check if 
(a) single-age bursts of star formation, 
(b) the stellar populations from the UV continuum fits, 
and (c) an additional BB component combined with the 
observationally-motivated stellar populations 
were capable of reproducing the emission lines
observed in these EELGs.

Our specific results from this work are:
\begin{enumerate}
    \item The stellar populations that dominate the continuum in 
    UV spectra are extremely young and metal-poor.
    We used the methods set up in \cite{chisholm19} to measure
    the age and metallicity of the stellar populations
    in J104457 and J141851 and found ages of 4.01 $\pm$ 1.13 Myr 
    and 3.37 $\pm$ 1.24 Myr respectively.
    The metallicities were measured at 0.09$\pm$0.05 Z$_\odot$ and 
    0.06$\pm$0.05 Z$_\odot$ for J104457 and J141851 respectively.
    We used both single-star and binary-star models and models with
    high-mass cutoffs at 300$M_\odot$ in addition to models with 
    high-mass cutoffs at 100$M_\odot$.
    The use of very massive star models did increase the 
    ionizing photon production rate but did not account for the
    increase needed to produce the extreme \ion{He}{2} we observe.
    
    \item Absorption features in the UV continuum also indicate
    young stellar populations.
    We found \ion{Fe}{5} and \ion{S}{5} absorption features in
    the UV continuum for both J104457 and J141851. 
    These features are indicative of young stellar populations
    in higher metallicity galaxies, but their observation in
    these low metallicity stellar populations confirms the
    presence of very hot stellar atmospheres which require
    young massive stars.
    
    \item We used photoionization modeling with \texttt{CLOUDY}
    to test whether single-age bursts of star-formation
    could reproduce the observed emission lines from our 
    EELGs.
    We find that \texttt{BPASS} single-age burst models
    match the emission lines from species in the low- and
    intermediate-ionization zones, but do not produce
    enough high-ionization photons to recreate the high-
    and very-high ionization lines, including \ion{O}{3}],
    [\ion{Ar}{4}], [\ion{Ne}{3}], \ion{He}{2}, and \ion{O}{4}.
    These models are frequently used to model the stellar 
    populations found in EELGs, but they are insufficient
    to reproduce the high-ionization emission lines from EELGs.
    
    \item We used the stellar populations fit in \autoref{sec:continuum}
    as inputs for our photoionization modelling with \texttt{CLOUDY}
    in an attempt to reproduce all emission lines with the
    stars observed in the galaxy.
    Although our observationally-motivated stellar
    models were able to reproduce the low-, intermediate-,
    and high-ionization lines, the very-high ionization lines
    like [\ion{Ne}{3}], \ion{He}{2}, and \ion{O}{4} were still
    out of reach for these models.
    The inability of the observed stars to reproduce the 
    observed emission indicates there is an additional source of 
    very-high energy ionizing photons produced in these galaxies.
    
    \item We added our observationally motivated stellar
    population from \autoref{sec:continuum} to a 
    BB of temperature 80kK that contributes 60\% and 
    70\% of the light from the young stars for J104457 and
    J141851 respectively.
    This ad hoc BB produces the very-high energy
    photons required to reproduce the very-high ionization
    lines like [\ion{Ne}{3}], \ion{He}{2}, and \ion{O}{4}.
    The combination of BB and observed stellar 
    population models is capable of reproducing the 
    entire suite of emission lines, from low-ionization
    [\ion{O}{2}] up to very-high ionization \ion{O}{4}.
    
    \item The \textit{high energy ionizing photon production problem}
    (HEIP$^3$) is often seen in EELGs where the 
    extreme emission lines like \ion{He}{2} and \ion{O}{4}
    cannot be reproduced by the stellar populations 
    expected to reside there or by other combinations
    of physical phenomena like collections of WR stars
    or ULXs.
    The ad hoc BB we used to model the
    high-energy photons needed to produce the 
    observed \ion{He}{2} solves the HEIP$^3$ for 
    the two EELGs studied in this paper by 
    simultaneously reproducing the suite of emission
    lines across all ionization zones from low to very-high.
    
    \item The BB we add to our 
    observationally-motivated stellar models could
    come from a number of physical sources, but we
    are unsure which sources would produce this
    set of ionizing photons.
    Some of the previously suggested solutions
    from the literature include shocks, WR stars,
    very massive stars, and ultra luminous X-ray 
    sources.
    We address these previous physical suggestions
    in \autoref{sec:whatbb}, but none of the 
    possible solutions stand out as a clear
    explanation of our high-energy photons.
    We require further observations at additional
    wavelengths to test more of these suggestions.
\end{enumerate}

We have listed numerous possibilities of physical sources that 
could produce the high-energy photons needed to explain the 
very-high ionization lines in J104457 and J141851.
Further exploration of these galaxies in the X-ray could help 
limit the options by exploring the range of photons at high-energies.
Additional information from spatially resolved \ion{He}{2} emission 
would provide information that could help determine if the source of 
the ionizing photons is centrally located or if it exhibits a larger 
spatial distribution.
Further UV spectroscopy could be useful in searching for \ion{O}{6} 
signatures related to some of the possible suggestions.
We leave these explorations of J104457 and J141851 to future work.

\begin{acknowledgments}
GMO is thankful for the support for this program HST-GO-15465,
that was provided by NASA through a grant from 
the Space Telescope Science Institute, which is operated by the 
Association of Universities for Research in Astronomy, Inc., 
under NASA contract NAS5-26555. DKE is supported by the 
US National Science Foundation (NSF) through Astronomy \& 
Astrophysics grant AST-1909198.

This paper made use of the modsIDL spectral data reduction reduction pipeline
developed in part with funds provided by NSF Grant AST-1108693 and a generous
gift from OSU Astronomy alumnus David G. Price through the Price Fellowship in
Astronomical Instrumentation. 

This work uses observations obtained with the Large Binocular Telescope (LBT).
The LBT is an international collaboration among institutions in the
United States, Italy and Germany. LBT Corporation partners are: The
University of Arizona on behalf of the Arizona Board of Regents;
Istituto Nazionale di Astrofisica, Italy; LBT Beteiligungsgesellschaft,
Germany, representing the Max-Planck Society, The Leibniz Institute for
Astrophysics Potsdam, and Heidelberg University; The Ohio State
University, representing OSU, University of Notre Dame, University of
Minnesota and University of Virginia.
\end{acknowledgments}

\noindent

\facilities{LBT (MODS), HST (COS)}

\software{\texttt{CLOUDY} (C17.01; \citealt{ferland17})}


\nocite{*}
\bibliography{extremeuv}

\clearpage

\end{document}